\documentclass{emulateapj}
\usepackage{natbib}
\newcommand{\Ha}{H$\alpha$}
\newcommand{\Msun}{$M_\sun$}
\newcommand{\HII}{\textsc{Hii}}
\newcommand{\HI}{\textsc{Hi}}
\newcommand{\mh}{\rm{[M/H]}}
\newcommand{\sfr}{$M_\sun$ yr$^{-1}$ kpc$^{-2}$}

\shorttitle{Radial SFH of NGC 300}
\shortauthors{Gogarten et al.}

\begin{document}

\title{The ACS Nearby Galaxy Survey Treasury V. 
  Radial Star Formation History of NGC 300}

\author{Stephanie M. Gogarten\altaffilmark{1},
  Julianne J. Dalcanton\altaffilmark{1},
  Benjamin F. Williams\altaffilmark{1}, 
  Rok Ro{\v s}kar\altaffilmark{1},
  Jon Holtzman\altaffilmark{2},
  Anil C. Seth\altaffilmark{3}, 
  Andrew Dolphin\altaffilmark{4}, 
  Daniel Weisz\altaffilmark{5}, 
  Andrew Cole\altaffilmark{6},
  Victor P. Debattista\altaffilmark{7},
  Karoline M. Gilbert\altaffilmark{1},
  Knut Olsen\altaffilmark{8}, 
  Evan Skillman\altaffilmark{5}, 
  Roelof S. de Jong\altaffilmark{9},
  Igor D. Karachentsev\altaffilmark{10},
  Thomas R. Quinn\altaffilmark{1}
}

\altaffiltext{1}{Department of Astronomy, University of Washington,
  Box 351580, Seattle, WA 98195; stephanie@astro.washington.edu}
\altaffiltext{2}{Department of Astronomy, New Mexico State University, Box
30001, 1320 Frenger St., Las Cruces, NM 88003}
\altaffiltext{3}{Harvard-Smithsonian Center for Astrophysics, 60
  Garden Street, Cambridge, MA 02138}
\altaffiltext{4}{Raytheon Company; 1151 E.\ Hermans Rd., Tucson, AZ 85756}
\altaffiltext{5}{Department of Astronomy, University of Minnesota, 116
  Church St. SE, Minneapolis, MN 55455}
\altaffiltext{6}{School of Mathematics and Physics, University of Tasmania, Hobart, Tasmania, Australia}
\altaffiltext{7}{RCUK Fellow at Jeremiah Horrocks Institute, University of Central Lancashire, Preston PR1 2HE, UK}
\altaffiltext{8}{National Optical Astronomy Observatory, 950
  N.\ Cherry Ave., Tucson, AZ 85719}
\altaffiltext{9}{Astrophysikalisches Institut Potsdam (AIP), An der Sternwarte 16, 14482 Potsdam, Germany}
\altaffiltext{10}{Special Astrophysical Observatory, N.Arkhyz, KChR, Russia}

\begin{abstract}
We present new Hubble Space Telescope (HST) observations of NGC 300
taken as part of the ACS Nearby Galaxy Survey Treasury (ANGST).
Individual stars are resolved in these images down to an absolute
magnitude of  $M_{F814W} = 1.0$ (below the red clump).  We
determine the star formation history of the galaxy in 6 radial bins by
comparing our observed color-magnitude diagrams
(CMDs) with synthetic CMDs based on theoretical isochrones.
We find that the stellar disk out to 5.4 kpc is primarily old, in
contrast with the outwardly similar galaxy M33.  
We determine the scale length as a function of age and find evidence
for inside-out growth of the stellar disk:
the scale length has increased from $1.1 \pm
0.1$ kpc 10 Gyr ago to $1.3 \pm 0.1$ kpc at present, indicating a
buildup in the fraction of young stars at larger radii.
As the scale length of M33 has recently been shown to have increased
much more dramatically with time, our results demonstrate that two galaxies with
similar sizes and morphologies can have very different histories.
With an $N$-body simulation of a galaxy designed to be similar to
NGC~300, we determine that the effects of radial migration should be
minimal.
We trace the metallicity gradient as a function of time and 
find a present day metallicity gradient consistent with that seen in
previous studies.
Consistent results are obtained from archival images covering the same
radial extent but differing in placement and filter combination.
\end{abstract}


\section{Introduction}

The standard model of galaxy formation has the inner parts of galaxies
forming before the outer parts, as a result of increasing timescales
for gas infall with radius
\citep{Larson1976,White1991,Burkert1992,Mo1998,Naab2006}.
This ``inside-out'' growth scenario has likewise been seen in
$N$-body/Smooth Particle Hydrodynamics simulations of disk galaxy evolution \citep{Brook2006},
and has been used to explain the observed abundance gradients in the Milky
Way via chemical evolution models 
\citep{Matteucci1989,Chiappini1997,Boissier1999}.

There are multiple mechanisms which could contribute to an observed
trend of star formation taking place more recently in the outer parts
of galaxies.  The first, as previously mentioned, is that gas does not
accumulate in the outer disk until later times, or that it accumulates
more slowly such that the total mass density of the outer disk increases
with time.  Alternatively, if the gas is in place but the star
formation timescale in the outer disk is longer, the \emph{stellar}
mass density would increase more slowly with time in the outer disk.

If disks do form ``inside-out,'' one would expect to see negative
radial gradients in age and metallicity.  Age gradients would result from a
higher percentage of older stars in the centers of galaxies, where
star formation started earlier.  This older population of stars would
enrich the surrounding gas, causing more recently formed stars toward
the center to be more metal-rich than stars formed in the outskirts of
the disk.

Negative gradients have been seen in several surveys of nearby
galaxies, and they appear to be common, if not universal, among larger
disks.
In the sample of nearby galaxies presented in \citet{Munoz-Mateos2007},
$\mathrm{FUV} - K$ color was used as a proxy for star formation rate
(SFR), since it
represents the ratio of young to old stars and shows what fraction of
the total star formation has been recent.
Large spirals were found to have predominantly negative gradients,
while low-luminosity
systems show a considerable scatter in the slopes of their gradients,
with both positive and negative values, indicating a wider range of
possible formation histories.
The \citet {MacArthur2009} study of nearby spirals using spectral
synthesis finds both
negative gradients and flat profiles in stellar age and metallicity.

A flat radial profile in age and metallicity may be consistent with
inside-out growth if gradients established early were
 erased by interactions or subsequent radial migration.
Especially in high-mass systems,
stars can scatter off of spiral structure and change their
radii while retaining circular orbits \citep{Sellwood2002}.
The simulations of \citet{Roskar2008a} showed that some percentage of
older stars currently residing in the outer disk
of a galaxy actually formed closer to the center of the disk and
migrated outward to their current location. 
Thus, an observed older
stellar population in the outer disk of a massive galaxy may not
necessarily indicate that the outer disk formed early.

The ideal galaxies for studying the evolution of disks in the absence
 of major mergers are undisturbed,
pure-disk (i.e., bulgeless) systems that maintain weak spiral structure to
suppress radial migration.
While one method of studying galaxy evolution is to compare
high-redshift galaxies with their local counterparts, difficulties
 with this method include the dramatic falloff of surface
brightness with redshift and the inability to trace the history of a
 single galaxy.
Color-magnitude diagrams (CMDs) derived from resolved stellar populations
can provide detailed, time-resolved studies of the disk evolution.
 This method restricts the possible targets to
relatively local galaxies. 

NGC~300, in the Sculptor Group, is the nearest isolated late-type disk galaxy,
and is thus an ideal target. 
\citet{Munoz-Mateos2007} have already suggested an
inside-out growth scenario for NGC~300 based on its broadband colors.
The resolution of the \emph{Hubble Space Telescope (HST)} enables us
to use CMD fitting to find the star formation histories (SFHs) of galaxies outside the Local Group.
The ACS Nearby Galaxy Survey Treasury \citep[ANGST;][]{Dalcanton2009}
was designed to create a
volume-limited sample of nearby galaxies, allowing for an unbiased
accounting of star formation in the nearby universe.  
The continuous radial strip of NGC~300 imaged as part of ANGST allows
us to recover the SFH of this galaxy out to $\sim5$
kpc.

\begin{deluxetable}{lcc}
\tablewidth{0pt}
\tablecaption{\label{tbl:properties}
Summary of NGC~300 and M33 properties}

\tablehead{
&
\colhead{NGC~300} &
\colhead{M33}
}

\startdata
Distance & 2.0 Mpc \tablenotemark{a} & 800 kpc \tablenotemark{b}\\
Type & SA(s)d \tablenotemark{c} & SA(s)cd \tablenotemark{c}\\
$M_B$ & -17.66 \tablenotemark{a} & -18.4 \tablenotemark{d} \\
Scale length ($K$) & 1.3 \tablenotemark{e} & 1.4 \tablenotemark{e} \\
Circular velocity & 97 km s$^{-1}$ \tablenotemark{f} & 130 km s$^{-1}$
\tablenotemark{g} \\
Estimated mass & $2.4\times 10^{10}$ \Msun \tablenotemark{f} &
$5\times 10^{10}$ \Msun \tablenotemark{g}
\enddata
\tablenotetext{a}{\citet{Dalcanton2009}}
\tablenotetext{b}{\citet{Williams2009a}}
\tablenotetext{c}{NED}
\tablenotetext{d}{\citet{Vila-Costas1992}}
\tablenotetext{e}{\citet{Munoz-Mateos2007}}
\tablenotetext{f}{\citet{Puche1990}}
\tablenotetext{g}{\citet{Corbelli2000}}
\end{deluxetable}

Evidence for inside-out growth (a decrease in mean stellar age with radius)
has also been found in M33, a galaxy very
similar to NGC~300 (see Table~\ref{tbl:properties} for a comparison
of their main properties).  
In M33, \citet{Williams2009a} and Holtzman et al.\ (in prep.) used
resolved stellar populations from four \emph{HST}/ACS fields to
derive SFHs using CMD fitting.  From these SFHs, they infer the
stellar surface density of the disk at different times throughout the
galaxy's history and find significant evolution in the scale length of
the disk.  The increase in scale length with time, indicating that the
SFR has been increasing in the outer disk, 
is suggestive of inside-out growth.

While outwardly similar, NGC~300 and M33 may have different histories.
M33 has a disk break at $\sim 6$ scale lengths \citep{Ferguson2007},
which is a common feature
in spiral galaxies \citep{Pohlen2006}.  However,
\citet{Bland-Hawthorn2005} showed that NGC~300 has a pure exponential
disk out to $\sim10$ scale lengths.  
There are environmental differences as well:
an \HI\ bridge between M33 and M31 is
suggestive of a history of interaction between these two galaxies
\citep{Braun2004,Bekki2008}, a finding that is confirmed
by the distribution of red giant branch (RGB) stars surrounding the two galaxies
\citep{McConnachie2009} and the evidence for tidal disruption of M33's
gas disk \citep{Putman2009}.
  In contrast, NGC~300 is fairly isolated on the
Sculptor filament, with only dwarf galaxies nearby \citep{Karachentsev2003}.
These differences may imply significantly different SFHs.

In \S\ref{sec:data}, we describe our data and reduction; in
\S\ref{sec:analysis}, we describe our methods for determining the star
formation history and present our results; we discuss their
implications for disk growth along with possible caveats and compare
our metallicity results to other observations in
\S\ref{sec:discussion}, and we conclude with
\S\ref{sec:conclusions}.  Archival data is presented in
Appendix~\ref{sec:archival}.
We adopt a WMAP
cosmology \citep{Spergel2007} for all conversions between time and redshift.

\section{Data and Photometry}
\label{sec:data}

\subsection{ACS Imaging}
\label{sec:imaging}

\begin{figure}
\plotone{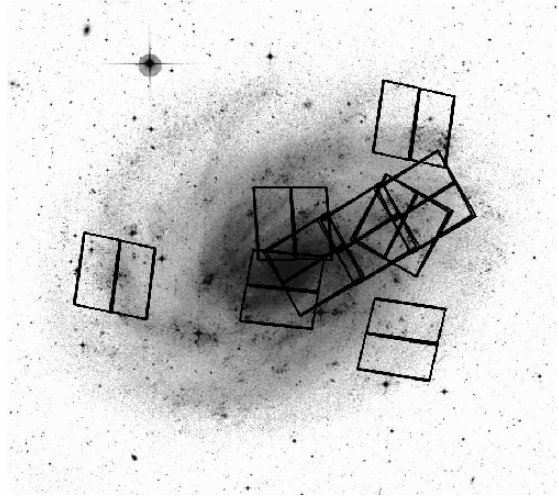}
\caption{\label{fig:overlays}
  DSS image of NGC~300 with ACS fields overlaid.
  The three contiguous ACS fields forming a radial strip
  are from ANGST, while the other fields are archival data.}
\end{figure}

\begin{figure*}
\plotone{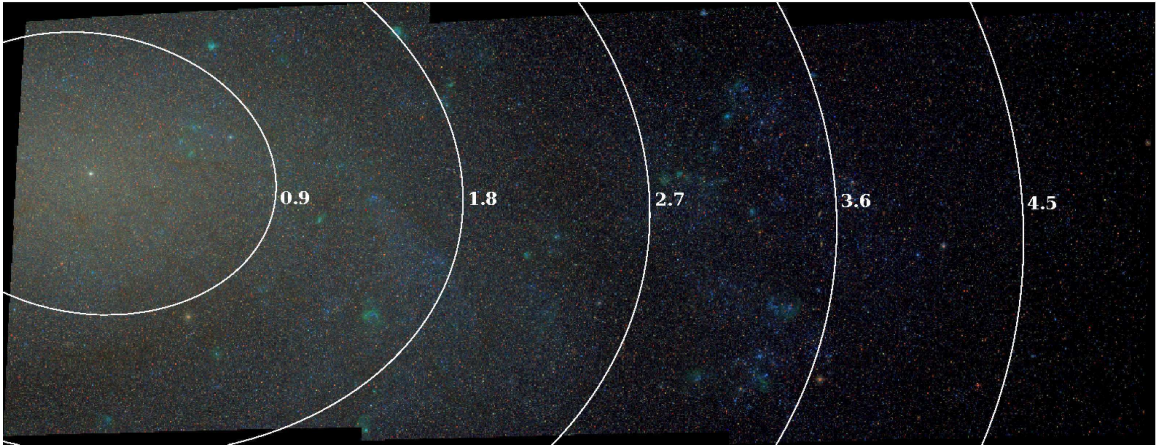}
\caption{\label{fig:color}
  Color image of the radial strip of NGC~300 observed as part of
  ANGST (red: $F814W$, green: $F606W$, blue: $F475W$).
  Inclination-corrected radial bins of width 0.9 kpc are overlaid as
  white lines, with radii in kpc labeled.}
\end{figure*}

As part of ANGST (GO-10915), we took \emph{HST}
  Advanced Camera for Surveys \citep[ACS;][]{Ford1998} observations of NGC~300
  during 2006 November 8-10.  
Three slightly overlapping fields were observed along a radial
strip from the center of the galaxy into the disk.  
Each field was observed for 1488s in $F475W$, 1515s in
$F606W$, and 1542s in $F814W$.
An additional deep outer field was planned as part of the ANGST
  program, which would have given more extended radial coverage of the
  disk; unfortunately, ACS failed before this observation was obtained.

An additional six archival ACS images, scattered across the disk of
NGC~300, are presented in Appendix~\ref{sec:archival}.
We include the archival data as a useful consistency check that the
ANGST data is representative of the disk as a whole.
  However, since the observing conditions were not identical (different exposure times,
  different filters, and no spatial continuity), we keep the analysis
  of the archival data separate from that of the new observations.
These images were originally taken as part of the Araucaria project to
determine Cepheid distances to nearby galaxies (GO-9492), so the fields were
selected to sample the Cepheid population at different galactocentric
distances and are therefore placed on more active star-forming regions
\citep{Bresolin2005,Rizzi2006}. 
The observations were obtained
between the dates of 2002 July 17 and December 25, with exposure times
of 1080s in $F435W$ and $F555W$, and 1440s in $F814W$.  
Each observation was split between two
exposures for cosmic-ray removal and coverage of the chip gap.
The exposures were calibrated and flat-fielded using the standard
\emph{HST} pipeline.
The locations of all ACS fields are shown in
Figure~\ref{fig:overlays}.

\subsection{Photometry}
\label{sec:photometry}

\begin{figure*}
\plotone{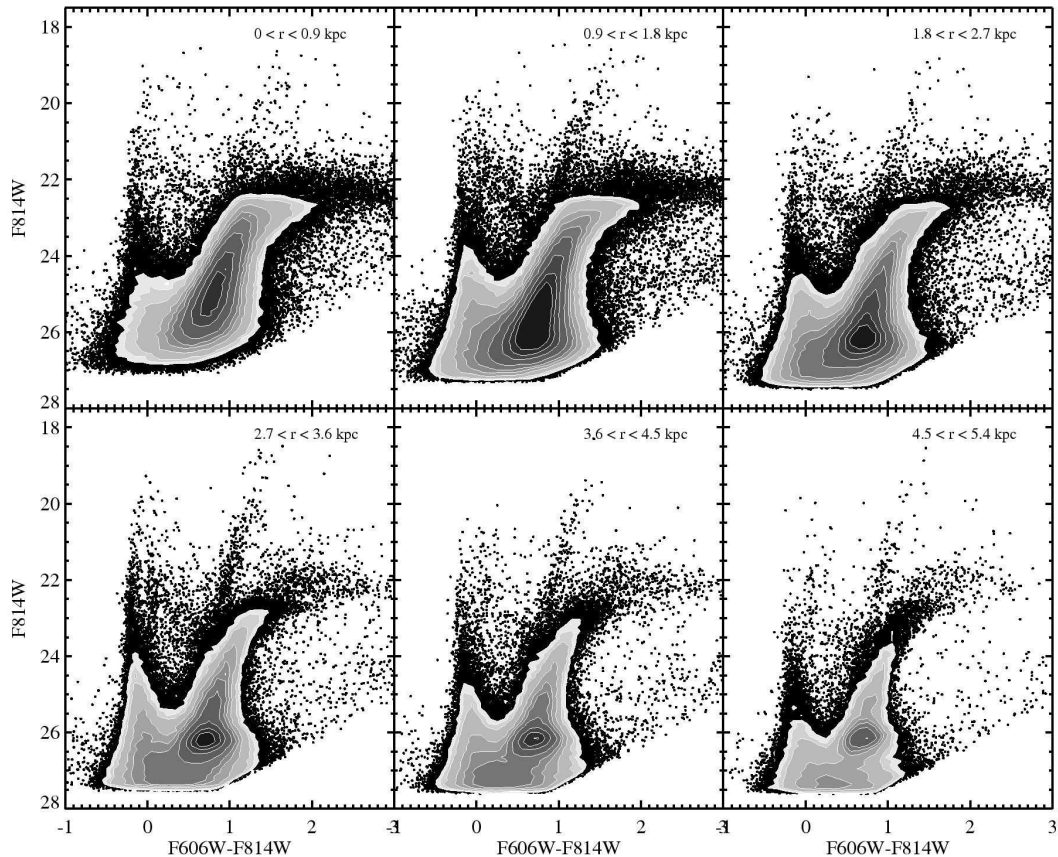}
\caption{\label{fig:cmds}
  CMDs of radial bins from ANGST data.  Only stars with S/N $>4$ and
  high-quality photometry have been included (see \S\ref{sec:photometry}).}
\end{figure*}

  While exposures were
combined using MultiDrizzle to produce images in this paper,
photometry was done using all individual exposures simultaneously.
For photometry, we use DOLPHOT, a modified version of
HSTphot \citep{Dolphin2000} optimized for the ACS.
DOLPHOT fits the ACS point spread function (PSF) to all of the stars
in each exposure, determines the aperture correction from the most
isolated stars, combines the results from all exposures, and converts
the count rates to the Vega magnitude system.  
Details of the photometry and quality cuts used for the ANGST sample
and archival data are given in \citet{Williams2009} and
\citet{Dalcanton2009}.
We require that stars in the
final sample are classified as stars, not flagged as unusable (too many bad or
saturated pixels or extending too far off the edge of the chip), have
$S/N > 4$, and have $|sharp_{F606W} + sharp_{F814W}| <0.274$.  
Sharpness indicates whether a star is too sharp (perhaps a cosmic ray)
or too broad, and these cuts
exclude non-stellar objects (such as background galaxies) that
escaped the earlier cuts. 
We also cut on the crowding
parameter, which is defined as how much brighter a star would
have been measured if nearby stars had not been fit simultaneously. 
Stars with a high crowding parameter are more likely to have erroneous
photometry, but a very strict cut has the effect of preferentially
removing young stars, since these are usually found in clusters.
We require $crowd_{F606W} + crowd_{F814W} <0.6$ mag.  

We characterize the
completeness of our sample in terms of magnitude, color, and
position by inserting at least $\sim 10^6$ artificial stars ($\sim50,000$
stars at a time) in each ACS field.
DOLPHOT is also used to perform the
artificial star tests.  Individual stars are inserted
into the original images and their photometry is re-measured.
Artificial stars are labeled as ``detected'' if they were found by
DOLPHOT and met the quality cuts described above.
The 50\% completeness limit for ANGST ranges from $F606W = 26.7, F814W
= 25.8$ in the crowded center of the galaxy to $F606W = 28.4, F814W =
27.4$ in the outer regions.  For the archival data, 50\% completeness
limits range from $F555W = 26.8$ to 27.7 and $F814W = 25.8$ to 27.2.

\subsection{Identification of Duplicate Stars}

The ANGST fields overlap by $\sim 7\arcsec$ so that
the fields could be aligned into a single radial
strip.  When combining the photometry of all stars into a single
catalog, 
we identify and remove duplicate stars in overlapping regions by binning star
positions in each field and finding bins that
contain stars from more than one field.  
The bin size is set at $1 \arcsec$, big enough so that nearly all bins
in the overlap regions contain stars in both fields but small enough
so that the edges of the overlap regions are fairly smooth.
The three ANGST fields are designated WIDE1, WIDE2, and WIDE3, with
WIDE3 containing the center of the galaxy and WIDE1 farthest from the
center.
For the
WIDE1-WIDE2 overlap, stars in WIDE1 are kept, and for the WIDE2-WIDE3
overlap, stars in WIDE2 are kept.
The World Coordinate System (WCS) for each of 3 fields is provided by the standard HST pipeline, but
there are very small offsets between fields.  We
found that relative to WIDE2, WIDE1 has an
offset of $\Delta \alpha = 0.23 \arcsec, \Delta \delta = 0.036
\arcsec$, and WIDE3 has an offset of
$\Delta \alpha = -0.13 \arcsec, \Delta \delta = -0.14
\arcsec$.  

\subsection{Dividing Stars into Radial Bins}

To assign an inclination-corrected
galactocentric distance to each star, we assume the following galaxy parameters:
$\alpha_0 = 13.722833\arcdeg$, $\delta_0 = -37.684389\arcdeg$ (galaxy center),
$i = 42\arcdeg$ (inclination), $\theta = 111\arcdeg$ (position angle)
\citep{Kim2004}.
To convert radius on the sky to physical distance, we assume a
distance of 2.0 Mpc to NGC~300, which is the distance we measure using
the tip of the red giant branch (TRGB) \citep{Dalcanton2009}.
The maximum radius we find for a star in the ANGST fields is $r =
5.35$ kpc.

Stars are divided into 6 radial bins based on their
(inclination-corrected) distance from the center of the galaxy.
Artificial stars are divided in the same way.
Our bins are spaced 0.9 kpc apart in radius, to encompass
all of the observed stars in annuli of equal width
(Figure~\ref{fig:color}).
The spacing of the bins is motivated by the length over which we
expect radial mixing to blur the populations \citep{Roskar2008a}.
The observed area is calculated for each bin so that we can derive the
correct surface density of star formation.
CMDs for stars in each bin are shown for the ANGST data in
Figure~\ref{fig:cmds}.

In the various fits to distance shown in \S\ref{sec:discussion} and
Appendix~\ref{sec:archival}, the distance used for each radial bin is the
mean distance from the center of pixels in that region.

\section{Star Formation History Analysis}
\label{sec:analysis}

\subsection{Method}

\begin{figure*}
\plotone{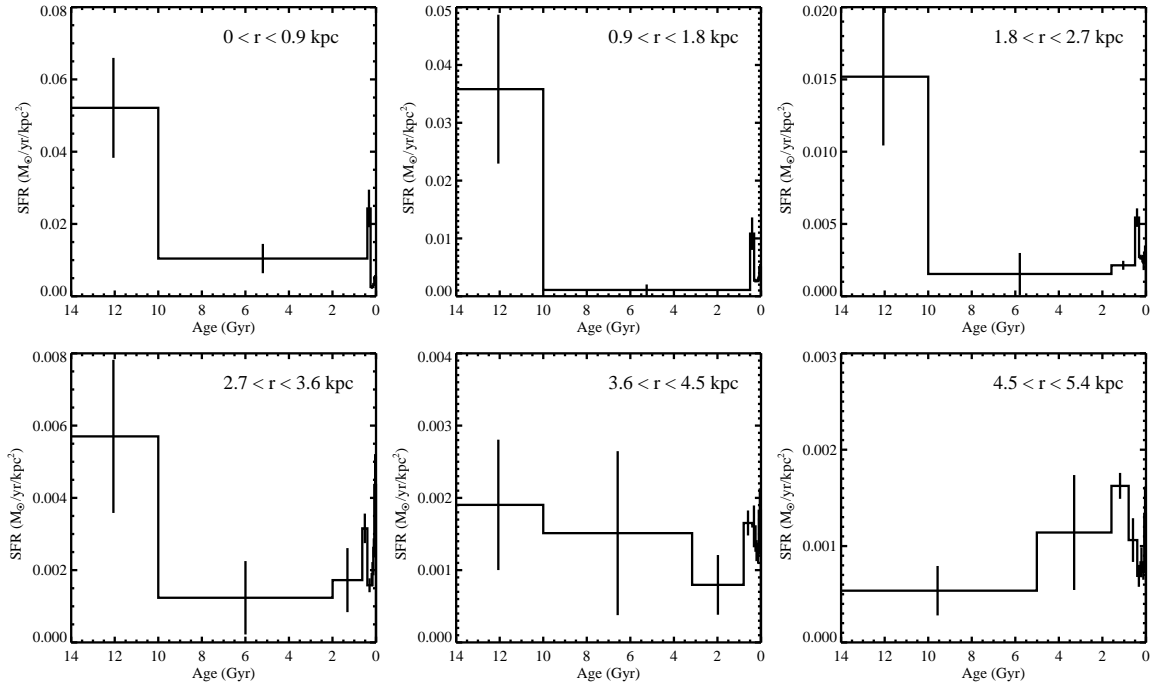}
\caption{\label{fig:sfh}
  Lifetime SFHs of radial bins from ANGST data.
  Error bars are the quadrature sum of the uncertainties from distance
  and extinction
  and the 68\% confidence interval from Monte Carlo
  simulations.
  SFRs have been scaled to a \citet{Kroupa2002} IMF.
The scale of the vertical axis differs in each panel.}
\end{figure*}

To derive the SFH of each radial bin, we use the well-established
technique of comparing the observed CMD to a set of model CMDs
\citep[e.g.,][]{Gallart1999,Hernandez1999,Holtzman1999,Dolphin2002,Skillman2003,Harris2004,Gallart2005}.
Typical fitting codes use stellar evolution models that
predict the properties of stars of different masses for a range of
ages and metallicities.  From the predicted luminosity and
temperature, the magnitudes of the stars are determined for a given
filter set.  Stars are then placed on a
synthetic CMD following the mass distribution of an assumed initial
mass function (IMF)
and binary fraction for each age and metallicity.
With distance
and extinction either set or included as additional free parameters,
these model CMDs are linearly combined until the best
fit to the observed CMD is found.  The ages and metallicities of the
CMDs that went into the best fit tell us the ages and metallicities of 
the underlying stellar
population, while the weights given to the CMDs provide the SFR at
each age.

We use MATCH, described in \citet{Dolphin2002}, to derive the SFH
for each radial bin.  This code finds the maximum-likelihood fit to
the CMD assuming Poisson-sampled data.
We assume an IMF with a slope of -2.35
\citep{Salpeter1955} between 0.1
and 120 \Msun\ and a binary fraction of 0.35. 
Given that 
our CMD only includes stars with masses $>1$ \Msun, adopting a single
Salpeter slope is likely to be a valid assumption.
While a power-law IMF was used by
MATCH for computational ease, the SFRs we show have been scaled
to a \citet{Kroupa2001} IMF.  This is possible because, for stars massive enough
to be observed, the two IMFs are very similar.
The choice of IMF affects only the normalization of the SFH, but not
the time dependence, so relative differences among age and radial bins
do not depend on the IMF.

Synthetic CMDs are constructed from the theoretical isochrones of
\citet{Girardi2002} and \citet{Marigo2008} for ages in the range 4 Myr--14 Gyr.
The isochrones younger than $\sim6\times 10^7$~yr
  were adopted from \citet{Bertelli1994}, with transformations to
  the ACS system from \citet{Girardi2008}.
Age bins are spaced logarithmically, since the
CMD changes much more rapidly at young ages than at old ages.
Metallicity is allowed to vary in the range $-2.3 < \mh < 0.1$, but is
not allowed to decrease with time.  See
\S\ref{sec:metallicity} for a further discussion of metallicity and
uncertainties therein.  

MATCH determines the best fit for distance and extinction by testing a
range of values.  The range for the distance modulus is 
 $26.3 \leq m-M \leq 26.7$, spanning the
range of values reported in the literature \citep[e.g.,][]{Butler2004,Sakai2004,Gieren2005,Rizzi2006,Dalcanton2009}
and the extinction range is $0.05 \leq A_V \leq
0.5$.  Additionally, up to 0.5 mag of differential extinction is
applied to young stars ($<100$ Myr), since these stars are more likely
to be found in dusty star formation regions \citep{Zaritsky1999,Zaritsky2002}.
The \citet{Schlegel1998} value for Galactic extinction is $A_V = 0.042$
in the line of sight to NGC~300,
but we expect the total value to be higher due to
local extinction within NGC~300 itself.  Additionally, we expect that
 dust content may vary across the extent of the disk, so extinction may
be different in each radial bin.

Completeness and observational errors are
accounted for by including the results of
artificial star tests.  
We identify the artificial stars that were placed
within each radial bin and supply MATCH with
their input and output magnitudes and whether they were detected
above the quality cuts of our photometry.
The density distribution of artificial stars mirrors the density of
detected stars within each bin, so MATCH accounts for any radial
variation in crowding.

We use the $V+I$ equivalent filter sets ($F606W+F814W$,
$F555W+F814W$) to
derive the SFHs in this paper, due to the greater depth of the
$F606W$ and $F555W$ data as compared to the $F457W$ and $F435W$ data.
As a consistency check, we also derived the SFHs using $F475W+F814W$ for the ANGST fields and
found SFHs that were
consistent within the error bars (which are larger in
$F475W+F814W$, especially in the inner regions), so our color choice
does not appear to affect our conclusions significantly.

We assess uncertainties
due to Poisson sampling of underpopulated regions in the CMD by running Monte Carlo simulations as
follows: for each region, we sample stars at random from the observed
CMDs until we reach the same number of stars as observed.
These
stars are then given as the input to MATCH with the distance and
extinction fixed, and the resulting SFH is
compared to the SFH from the original data.  We repeat this process 
100 times and define our sampling error as the values which encompass
68\% ($1\sigma$) of the Monte Carlo tests.  Our final error bars are
the quadrature sum of this error and the systematic errors from
fitting the distance and extinction.
Our error bars do not include systematic uncertainties in the stellar
evolution models.

\subsection{Distance, Extinction, and Crowding Effects}
\label{sec:effects}

\begin{figure*}
\plotone{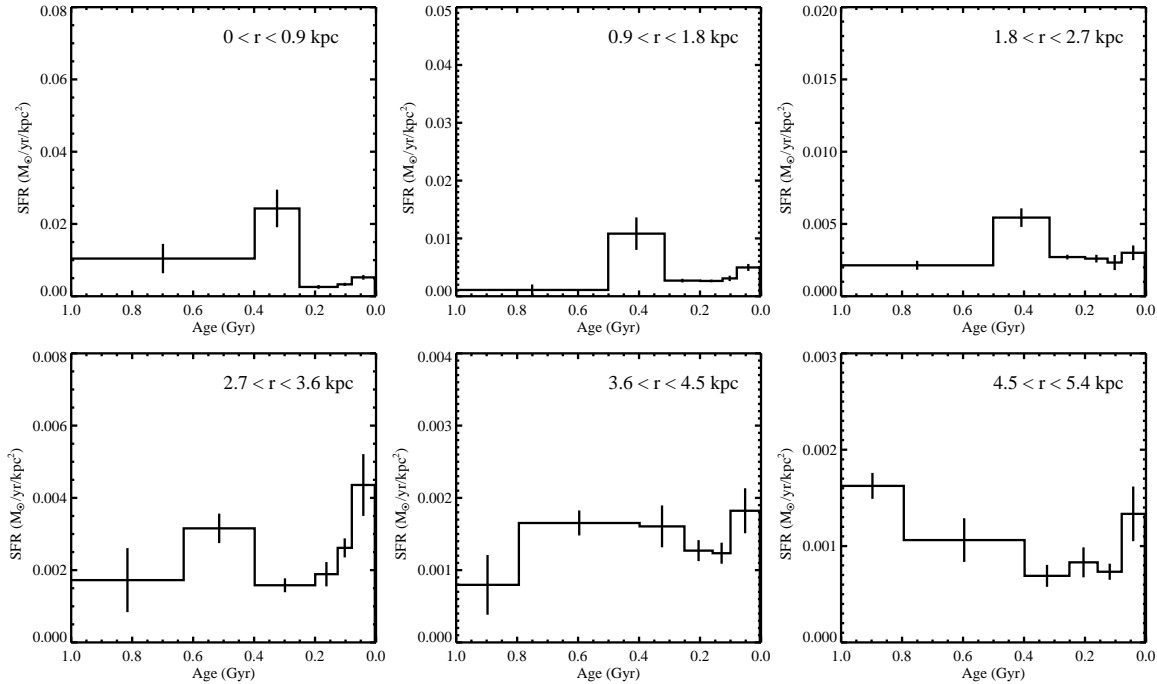}
\caption{\label{fig:sfh_recent}
  Recent ($<1$ Gyr) SFHs of radial bins from ANGST data.
  Errors are calculated as in Figure~\ref{fig:sfh}.
  SFRs have been scaled to a \citet{Kroupa2002} IMF.
The scale of the vertical axis differs in each panel.
  The bin at $2.7<r<3.6$ kpc contains a prominent spiral arm.}
\end{figure*}

Because our data span a large range of stellar densities, the
effective completeness varies strongly with radius.  To deal with this
variation, we pursued two independent methods.
The first approach is to consider only the portion of the CMD that
is 50\% complete for each radial bin, so that larger portions of the CMD can
be used in the outer, less crowded regions.  The second approach is to
find the completeness in the most crowded region (the center of the
galaxy) and apply that magnitude cut to
every region uniformly.  This
removes useful data in the outer regions, but
ensures that variations we see in the SFH across the disk are not due
to variations in photometric depth.  
We have employed both methods and compared the results.  

When we allow the completeness cut to vary with radius, the photometry in
the outer four radial bins is complete to well below the red clump
(see Figure~\ref{fig:cmds}).  This depth allows
for a more consistent distance measurement across these four fields; 
derived distance modulus values for the ANGST data in the six radial
bins were as
follows: $m-M = 26.61 \pm 0.07, 26.60 \pm 0.07, 26.43 \pm 0.09, 26.43
\pm 0.09, 26.42 \pm 0.09, 26.43 \pm 0.09$.
All of the four outer bins, with depths that resolve the red clump, produce a consistent estimate for the
distance of $m-M = 26.43 \pm 0.09$.  The inner fields, where crowding
limits the depth, have a
discrepant distance.  
We hereafter fix the distance to $m-M=26.43$ in all fields.
Even if this distance is not absolutely correct, what is important is
that setting the distance to the same value for all fields ensures
that any \emph{relative} changes in the positions of stars on the CMDs
between radial bins are interpreted as differences in SFR and/or
extinction.

Comparing the SFH when depth is allowed to increase with radius and
the SFH where depth is at a fixed magnitude throughout,  the recent SFHs for ages $<1$ Gyr are very similar, probably because the recent SFH is constrained by younger,
brighter stars and is not significantly affected by the elimination of
stars near the bottom of the CMD. 
For ages $> 1$ Gyr, the most significant difference is a 40--80\%
increase in the
SFR at 8--14 Gyr for the outer 3 radial bins when restricted to a
uniform shallow depth.  There is a corresponding overall
decrease in the more recent SFR (1--8 Gyr).
These tests indicate that using these time bins, the central region's
SFH may be more
biased to older ages than would be measured with less
crowded data.

To address this effect, we determined bin
widths that reflect
our sensitivity to age as described in \citet{Williams2009b}.
The resulting time bins are wider at intermediate ages (1--10 Gyr) for
the most crowded
regions, and thus robust against uncertainties in SFR within this age range.
Since the most recent time bins contain small numbers of stars,
assessing the statistical significance of these bins is more
difficult, and thus we present the past $\sim 80$ Myr as a single time
bin in all regions.  Changes in SFR on shorter timescales do not
affect any of the conclusions of this paper.

Given that using all available data allows us to better constrain the
SFH in the outer regions,
the SFHs presented in this paper are those derived allowing the 50\%
completeness limit to vary with radius.

The mean extinction values found for the ANGST data are $A_V = 0.10
\pm 0.05$ for all bins.  
Note that these values do not include the 0.5 mag of differential
extinction applied only to young stars ($<100$ Myr).
We discuss additional tests applying differential extinction to older
stars as well in \S\ref{sec:extinction}.

\subsection{Radially Resolved Star Formation Histories}
\label{sec:results}

The SFH for each radial bin, as derived by MATCH, is shown in
Figures~\ref{fig:sfh} and \ref{fig:sfh_recent}; the former shows the
lifetime SFH, and the latter focuses on the recent SFH ($< 1$ Gyr). 
The overall behavior
of SFR vs.\ age changes with radius; central regions have a higher
percentage of
old stars, and the percentage of young stars increases with radius.
The bulk of a spiral arm, visible on the color image in Figure~\ref{fig:color}, is
within the radial bin at $2.7 < r < 3.6$ kpc.  This region also has the most
dramatic increase of SFR in the most recent time bin as compared with
its level in the past Gyr, as can be seen
in Figure~\ref{fig:sfh_recent}.

Appendix~\ref{sec:archival} presents the SFH for archival data in
identical radial bins.  Rather than a continuous radial strip, these
fields are scattered throughout the disk.  The general agreement
between the radial SFHs of fields selected in two different ways
illustrates that we are not biasing the results by looking at only a
portion of the entire disk.

We present the full SFH for all radial bins, including metallicities,
in Table~\ref{tbl:10915}.  Metallicity will be discussed further in
\S\ref{sec:metallicity}.

\section{Discussion}
\label{sec:discussion}

\subsection{Stellar Disk Growth}

\begin{figure}
\plotone{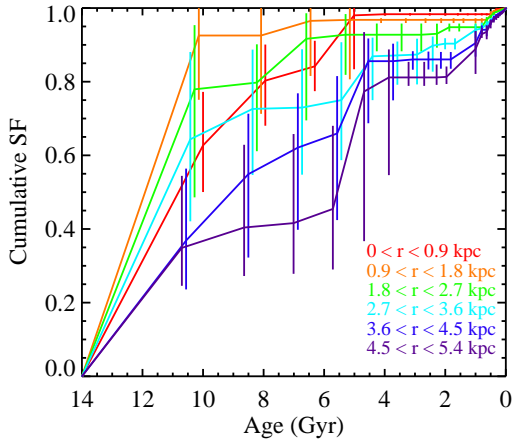}
\caption{\label{fig:cumul}
  Cumulative star formation (i.e., fraction of stars formed vs.\ age
  of galaxy) for all radial bins for ANGST data.
  Bin edges have been offset to avoid overlapping error bars.}
\end{figure}

We can use the SFH to infer the past structure of the disk.  
There is evidence for ``inside-out'' growth (Figure~\ref{fig:sfh}),
as early star formation was more prominent in the inner disk.
These trends are quantified in
Figure~\ref{fig:cumul}, where we plot the cumulative SFH. 
Cumulative plots are presented at the full resolution of the CMD fit,
since changes in SFR due to uncertainties typically occur in adjacent
time bins and thus have a very small effect on the cumulative SFH.
The outer parts of the disk formed a greater fraction of
their stars at recent times than the inner parts of the disk,
consistent with the ``inside-out'' growth scenario.  However, even the
outer regions are fairly old, with $\sim75$\% of stars formed by 4 Gyr
ago.

We emphasize that
when we refer to our ``outer'' disk observations in this
paper, the galactocentric distances are still within 5.4 kpc for a
galaxy that has been shown to extend to at least 14 kpc
\citep{Bland-Hawthorn2005}.  Therefore, the reader should keep in mind
that all distinctions between ``inner'' and ``outer'' we can make
with our data are still within what some might consider the ``inner''
regions of NGC~300.

\begin{figure}
\plotone{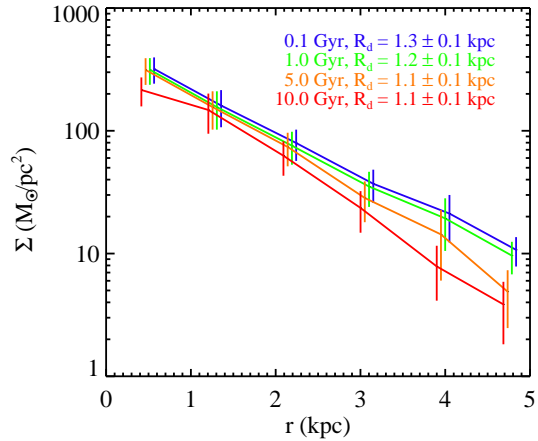}
\caption{\label{fig:surfdens}
  Stellar surface density versus radius for multiple time bins.  Points have
  been offset from one another by 0.05 kpc to avoid overlapping error bars.
  To compare this surface density
  with one assuming a \citet{Kroupa2001} IMF, one should divide by 1.5.}
\end{figure}

We can also infer past stellar surface density if we assume that the
majority of stars formed within the radial bin in which they are
currently found (see \S\ref{sec:migration}).
We calculate the stellar mass in each radial bin that formed
before a given time by summing all the mass formed up to that time in the
derived SFH.
For the purposes of calculating stellar surface density we require a uniform
set of time bins for all observed regions, so we choose a coarse
binning scheme of 4--100 Myr, 100 Myr--1 Gyr, 1--5 Gyr, 5--10 Gyr, and
10--14 Gyr.  
The resulting surface mass density profile is shown in
Figure~\ref{fig:surfdens}.
We omit the most recent time bin from the plot, as the surface
density has not
changed significantly in the past 100 Myr.
The stellar surface density has changed more substantially in the outer
regions of the disk than in the central regions.

For each time bin, we fit an exponential disk model to derive the
scale length of the disk; the results of these fits are shown in Table~\ref{tbl:fits}.  
The scale length increased slightly
over the galaxy's history, from $1.1 \pm 0.1$ kpc at early times to
$1.3 \pm 0.1$ kpc
by 1 Gyr ago.  However, our error bars are also consistent with no scale
length evolution.
A predominantly old disk is in contrast to the dramatic changes in
scale length than has been seen in
M33 or has been inferred from in situ studies of disk evolution at
high redshift \citep[e.g.,][]{Trujillo2005,Barden2005,Azzollini2008}.
M33 has
been shown to have a scale length that increases by nearly a factor of
2 inside the disk break \citep{Williams2009a}, in agreement with the
predictions of \citet{Mo1998}.  
Although M33 is a near twin of NGC~300 in mass and morphology, 
NGC~300 is different in that it lacks a disk
break in its exponential profile \citep{Bland-Hawthorn2005}
and is much more isolated than M33.

The value for the scale length derived by summing over the
stellar mass formed in the derived SFH agrees remarkably well with the
scale length for the $K$-band stellar mass surface density of
$1.29^{+0.02}_{-0.03}$ \citep{Munoz-Mateos2007}.  
Since massive stars have a high mass-to-light ratio in the $K$-band,
agreement with the scale length as traced by the stellar mass
formed in the disk is expected for a galaxy dominated by old stars
($\gtrsim 5$ Gyr).
Scale lengths measured
at shorter wavelengths are predictably larger, since these trace
younger stellar populations: 1.47 kpc in $I$ \citep{Kim2004} and 2.17
kpc in $B_J$ \citep[][scaled to a distance of 2.0 Mpc]{Carignan1985}.
M33 has a similar scale length in the $K$-band of 1.4
\citep{Munoz-Mateos2007}.
In the outer fields of M33 (4--6 kpc, or $\sim3$--4 scale
lengths), only $\sim20$\% of the stars formed by
8 Gyr ($z\sim1$), whereas in NGC~300, 50-70\% of stars in the 
equivalent outer bins had formed by this time.
Thus, the disk of NGC~300 appears to be older overall
than M33.  
If NGC~300 experienced significant inside-out growth, it may have
happened earlier than we are sensitive to with CMD fitting.

\begin{figure}
\plotone{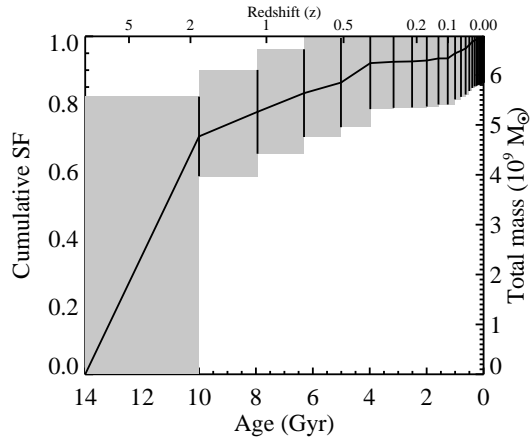}
\caption{\label{fig:cumul_global}
  Cumulative star formation for the entire galaxy out to 5.4 kpc.
  Shaded region indicates uncertainty.}
\end{figure}

Figure~\ref{fig:cumul_global} shows the cumulative SFH for all of
NGC~300 out to 5.4 kpc.  For each radial bin, we assumed the observed
SFH was characteristic of the entire annulus, and
summed stellar mass formed over all radial bins.  Overall, $\sim75$\%
of stars in this portion of NGC~300 formed by 8 Gyr ago.

\subsection{Complicating Effects}

\subsubsection{Photometric Depth}
\label{sec:depth}

\begin{figure*}
\plotone{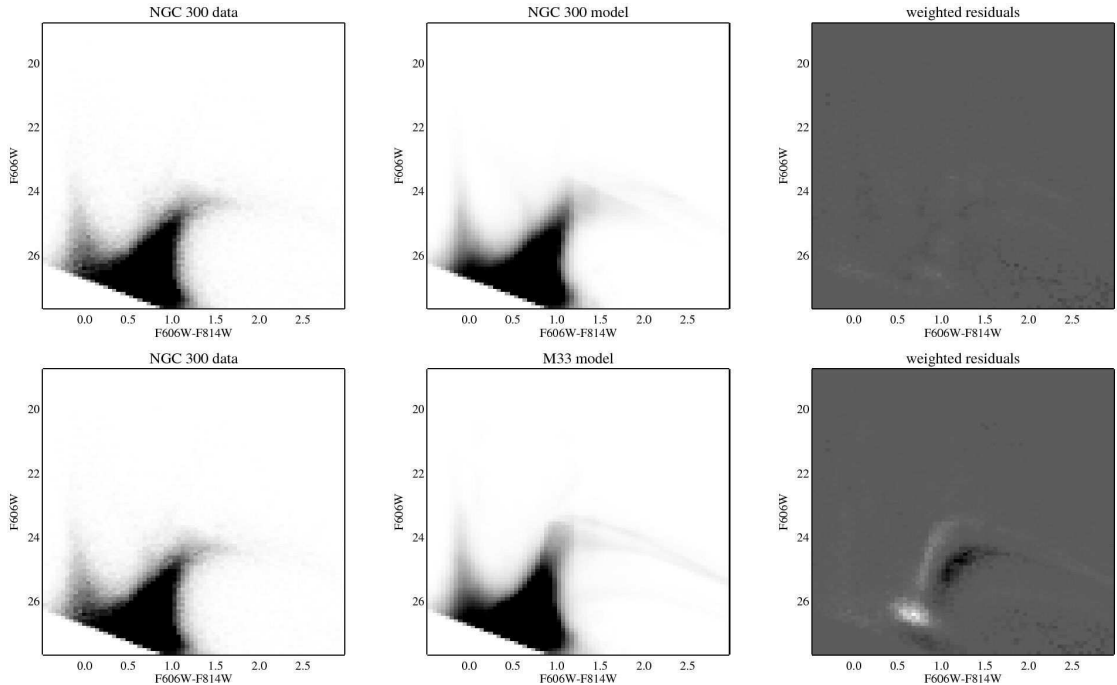}
\caption{\label{fig:m33resid}
  Top panels: results of fitting model CMDs to the NGC~300 data.
  Left to right: NGC~300 binned CMD, best fit CMD to the NGC~300
  data as found by MATCH, weighted residuals ((data - model)/Poisson noise).
  Bottom panels: results of fitting a synthetic CMD created from the SFH of the
  M33 field at $\sim2.5$ kpc to
  the NGC~300 CMD at 1.8--2.7 kpc.  
  Left to right: NGC~300 binned CMD, best fit CMD to the M33 data,
  weighted residuals.
  The grayscale is defined as
  (NGC~300 data) - (M33 model)/(Poisson noise), with darker areas
  being higher values
  (i.e. black areas show where there are more stars in NGC~300 and
  white areas show where there are more stars in M33).  NGC~300 has a
  redder RGB and a fainter red clump, which is consistent
  with an older, more metal-rich stellar population.}
\end{figure*}

Since the red clump is not resolved in the crowded inner regions,
the central SFH (older than $\sim1$ Gyr) is less certain (see \S\ref{sec:effects}), and could possibly
affect our scale length derivation. 
As an alternative, 
we fit an exponential profile to only the outer four radial bins, and
found essentially the same results as the fit to all the regions,
albeit with larger error bars.  Thus the crowded inner regions are not significantly
affecting the scale lengths we derive.

To further investigate whether the SFH found for M33 is consistent with the
observed stellar populations of NGC~300, we constructed model CMDs
from the M33 SFHs \citep{Williams2009a} using the photometry
statistics of our NGC~300 data.
For each
radial bin in NGC~300, we used the SFH from the observed field in M33 that most
closely matched in radius.  
We ran MATCH on the NGC~300 data while
enforcing the M33 SFH model, scaling the SFH overall so that the
number of stars produced by the model was equivalent to the number of stars
observed in NGC~300.  
In all cases, there were
substantial residuals and an obvious mismatch between the NGC~300 data and
M33 SFH model.  

Figure~\ref{fig:m33resid} compares the original MATCH fit to the
NGC~300 $1.8<r<2.7$ kpc radial bin with the results of fitting the M33 model
for 2.5 kpc to the same data.
The top series of panels shows the MATCH fit to the NGC~300 data,
along with the weighted residuals (data minus model divided by Poisson
noise).  Residuals for the best fitting model are very small.
The bottom series of panels shows the M33 model CMD at this
radius, along with the (very substantial) weighted residuals. 
We present this radial bin as an example; residuals were comparably
large in other bins.   
Figure 8 in \citet{Williams2009} demonstrates the effect of age and
metallicity on the red clump and asymptotic giant branch (AGB) bump:
higher metallicity gives redder colors and fainter magnitudes, while
older ages result in an even stronger push towards fainter magnitudes.
As shown in Figure~\ref{fig:m33resid}, the
RGB is redder in NGC~300 than in M33, and the red clump
is fainter, consistent with
an older, more metal-rich stellar population.  
This comparison confirms that the NGC~300 data is not consistent with
an M33-like SFH, but rather that NGC~300 formed the majority of its
stars much earlier than M33.

\subsubsection{Extinction}
\label{sec:extinction}

Given what appear to be dusty regions visible toward the center of
NGC~300 (Figure~\ref{fig:color}), it is possible that differential
extinction may be a factor for all ages of stars.  
Our SFH derivation accounts for
differential extinction of young stars ($<100$ Myr), but if there are
dust lanes present, differential extinction could affect older stellar
populations as well.
\citet{Roussel2005} studied extinction in \HII\ regions in NGC~300 and
found values for A(\Ha) ranging from 0.15 to 1.06.  While we expect
the \HII\ regions to contain predominantly young stars, we can use
their estimates as a rough guide for assessing extinction.  In MATCH,
the differential extinctions for young stars and for the stellar
population in general are set by two different parameters, with the
total extinction for young stars randomly assigned up to the sum of
these two values.  Given that we set the maximum extinction for young
stars to be 0.5, if we additionally supply a value for all stars of
0.5, the young stars would receive extinction roughly in the range
observed by \citet{Roussel2005}.

We experimented with adding extinction parameters from
0.1 to 0.5 mag, meaning that each star is randomly assigned an
extinction from zero to an upper limit of the extinction parameter. 
We found that the fit quality decreased with the
amount of differential extinction added.  In the central bin, where
the presence of dust is clearly visible, this difference was minimal:
up to 4\%.  In the remaining bins, the fit was up to 25--40\% worse.
For the central bin, the next best fit after no additional
differential extinction is when this parameter is set at 0.2
magnitudes, so we recalculated the surface density at each time bin
using up to 0.2 mag of additional extinction for only the central
bin.
Within the error bars, the scale length evolution remained the same.

\subsubsection{Radial Stellar Migration}
\label{sec:migration}

\begin{figure*}
\plotone{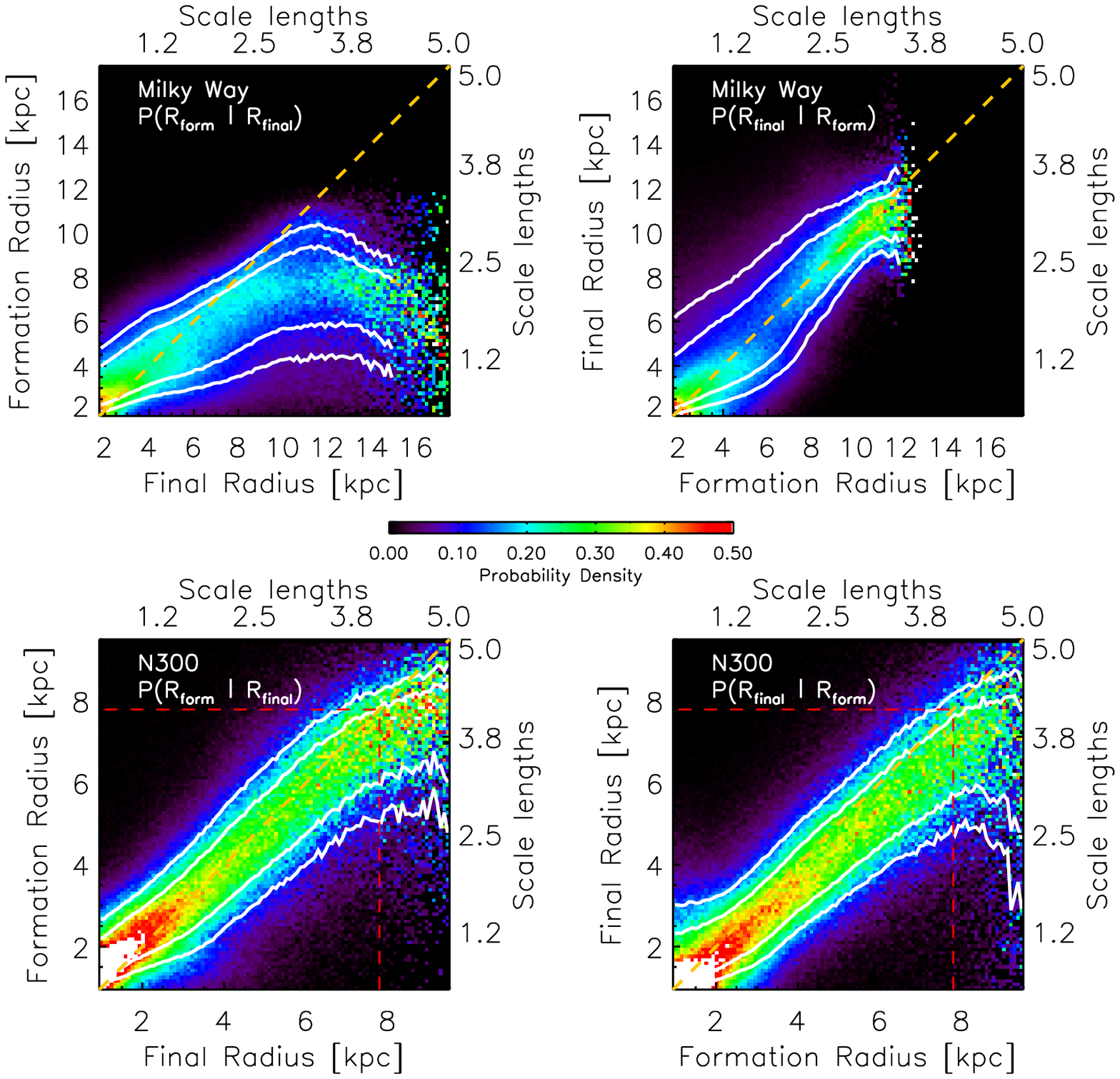}
\caption{\label{fig:nbody}
Probability of migration, for MW model (top) and N300 model (bottom).
Left-hand panels show, for a star at a given $R_{final}$, the
probability that it formed at $R_{form}$.  Right-hand panels show, for
a star at a given $R_{form}$, the probability that it will end up at
$R_{final}$.  
Contours show 50\% and 75\% probability.  The red dashed lines in the
N300 panels show the radial extent (in scale lengths) of NGC~300 studied in the current
work.
Yellow dashed lines show one-to-one correspondence between formation
and final radii, i.e. the expected position with no migration.
The N300 model shows less migration than the MW model.
}
\end{figure*}

\citet{Wielen1977} suggested that substantial migration among the stars
      in a disk galaxy
      is expected, and \citet{Sellwood2002} showed that transient
      spirals can provide an efficient mechanism to redistribute stars radially.
Recently, \citet{Roskar2008a} have reproduced
      this phenomenon in  
      simulations of growing disks.  Therefore, the radius at which
      stars are observed  
      may not be the radius at which they formed \citep[e.g.,][]{Wielen1996}.  
According to \citet{Roskar2008a}, migration effects become 
more important with increasing radius, in terms of the fraction of stellar mass that is 
composed of migrated vs.\ in-situ stars. In their Milky Way-analogue model, 
migration significantly influenced the surface density beyond 
$\sim 2$ scale lengths. This suggests that the
three outer radial bins in NGC~300 may contain stars that have
migrated from nearer the center of the galaxy.
Migration of stars from the inner to the outer disk could 
potentially erase the signature of inside-out growth in NGC~300,
because radial bins could be contaminated by stars that did not
actually form within them.
If NGC~300 has had more substantial migration than M33, that could
explain the discrepancy between the scale length evolution of the two galaxies.
On the other hand, both M33 and NGC~300 are significantly less massive than the disks used in the
\citet{Roskar2008a} simulations, so the effect of radial mixing may be less severe, due to
weaker spiral structure.

To examine the effects of migration in NGC~300, we ran an $N$-body
simulation of a disk galaxy designed to be similar in
mass and angular momentum to NGC~300.
The mass at 11.8 kpc (assuming a distance of 2.0 Mpc to NGC~300) was
measured to be $2.4 \times 10^{10}$ \Msun\ by \citet{Puche1990} using velocity
measurements from \Ha.
However, the disk of NGC~300 actually extends significantly farther
than this \citep[e.g.,][]{Bland-Hawthorn2005}.  We estimate the total
mass of NGC~300, including baryonic and non-baryonic mass, to be
$\sim$$10^{11}$ \Msun\ for the purposes of the simulation.

To estimate the spin parameter $\lambda$, we use
the formula derived by \citet{Hernandez2006}:
\begin{equation}
\label{eqn:lambda}
\lambda = 21.8 \frac{R_d/{\rm kpc}}{(V_d/{\rm km\ s}^{-1})^{3/2}},
\end{equation}
where $R_d$ is the disk scale length and $V_d$ is the rotation
velocity of a flat rotation curve.
The $I$-band scale length of NGC 300 was measured by \citet{Kim2004} as $R_d =
1.47$ kpc, and the velocity of  $V_d
= 93.4 \pm 8.2$ km s$^{-1}$ was measured by \citet{Puche1990} with \HI\
observations.  Putting these
values in equation~\ref{eqn:lambda}, we find that for NGC~300, $\lambda
\approx 0.0355$.
The Milky Way model (MW) uses a total mass of $10^{12}$ \Msun\ and a spin
of $\lambda \approx 0.039$.
Both models are evolved to 10 Gyr;
the scale lengths of the two models at the end of 10 Gyr are 1.9 kpc and 3.5 kpc
for N300 and MW respectively.

Comparing the results of the simulated NGC~300 with the Milky
Way-sized galaxy from \citet{Roskar2008a}, we find that the effect of
radial migration is substantially reduced in the smaller galaxy.
In Figure~\ref{fig:nbody}, we show the probabilities of migration for
the Milky Way model (top) and the NGC~300 model (bottom).
The left-hand panels should be interpreted as, ``if a star is
currently at $R_{final}$, what is the probability that it formed at
$R_{form}$?''  The right-hand panels, conversely, show ``if a star forms
at $R_{form}$, what is the probability that it will end up at
$R_{final}$?''  In all panels, probability is indicated by color, with
redder colors meaning higher probabilities.  
In the MW model, most stars form at $<12$ kpc (top right
panel), but these stars are found at radii up to 16 kpc (top left
panel).
Stars at all radii have a tendency to end up farther out than
where they formed.
In the N300 model, however, most stars are found
near their birth radii.

Figure~\ref{fig:nbody} shows that particles in the N300 model undergo much less
radial migration than those in the more massive MW model.
Since migration occurs when stars scatter off disk asymmetries such as
spiral arms \citep{Sellwood2002}, weaker spiral structure reduces the
probability of migration.  
A  disk
must remain kinematically cool to sustain recurrent transient spirals
\citep{Sellwood1984}, and the only way to cool a stellar disk is by star
formation, which repopulates circular orbits with young stars. 
Hence, if the majority of a disk is old, as in NGC~300, we can expect
that it experienced radial redistribution early in its evolution,
but not in recent years. 
Because
our time resolution at old times is poor, we would not be sensitive
to such early evolution.

In contrast to NGC~300, the majority of stars in M33 formed more recently
\citep{Williams2009a}, so M33 may have experienced more migration in recent times.
M33's interaction with M31 may also have driven enhanced spiral
structure, which in turn drives more migration
\citep[e.g,][]{Quillen2009}.  NGC~300 is in a much
more isolated environment.  
These factors combined suggest that one would expect M33 to have
experienced more migration than NGC~300, not less.

\begin{figure*}
\plotone{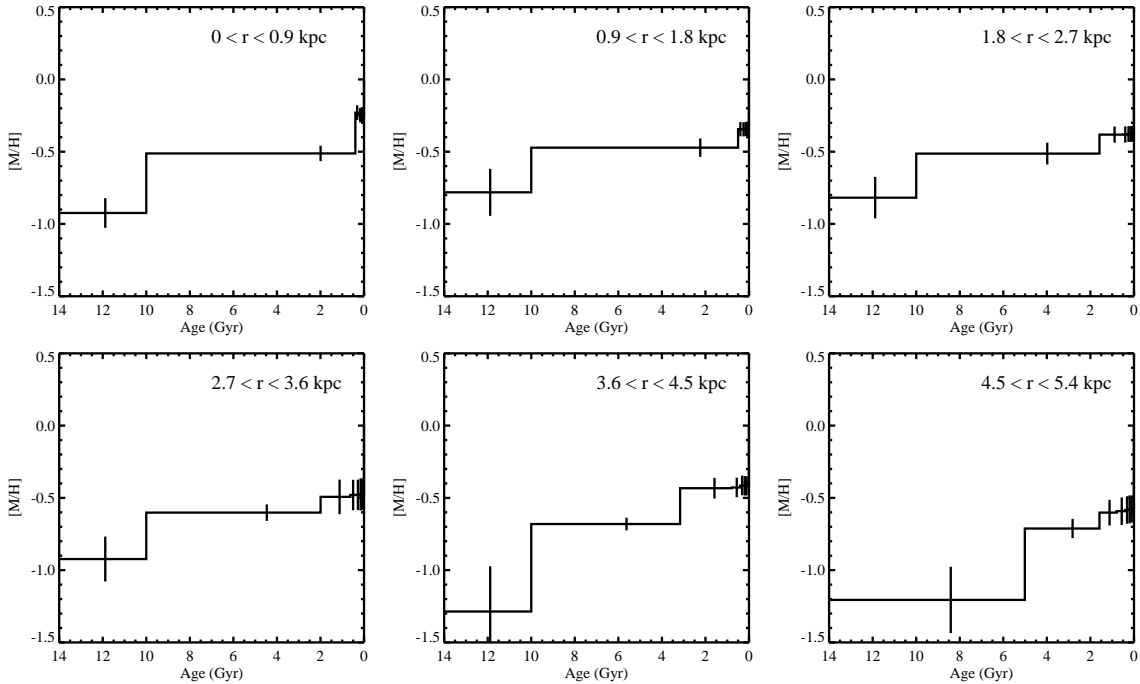}
\caption{\label{fig:metal}
  Mean metallicities corresponding to SFHs in Figure~\ref{fig:sfh}.
  Errors are calculated as in Figure~\ref{fig:sfh}.}
\end{figure*}

In summary, while migration will tend to erase the signature of disk
growth, it is not the only way to obtain scale lengths which do not
evolve as a function of age of stellar population.  From our
modeling, we can surmise that mass is an important factor in
the extent of radial migration, with less massive systems experiencing
less redistribution. Consequently, the lack of scale length evolution
in NGC~300 is attributable to a factor other than radial migration, in
this case most likely the rapid early growth of the disk.

Further evidence that NGC~300 has not experienced much migration comes
from the kinematics of the globular cluster systems.
\citet{Olsen2004} and \citet{Nantais2008} found that the globular
clusters in NGC~300 had kinematics matching that of the \HI\ disk,
indicating that their present location is where they formed.  
Since both heating and migration are caused by spiral arms, this
finding would suggest a lack of migration as well.
In a more massive galaxy, NGC~253, asymmetric drift of the globular
clusters indicates stronger radial diffusion.  This result agrees with
the prediction of the models described here in suggesting that more
massive galaxies experience greater migration.  The old globular
clusters with disk kinematics in NGC~300 also fits with the
interpretation that the majority of stars formed early.

\subsubsection{Comparison with M33's scale length evolution}

We considered whether the NGC~300 data is consistent with the stronger
scale length evolution seen in M33, such that the observed weaker
evolution is due solely to the shallower depth of the NGC~300 CMDs.
In other words, if M33 were at the distance
of NGC~300, would the results of \citet{Williams2009a} be recoverable?  
Unlike the test in \S\ref{sec:depth}, in which we saw that the SFH of
M33 is not compatible with the NGC~300 data, here we are looking at
the effect of depth on the M33 data.

We calculated photometry
of the M33 data using only one 1300s exposure in $F606W$ and $F814W$
for each of the ACS fields analyzed in \citet{Williams2009a}.
When deriving the SFH
we considered only the portion of the CMD that was 2 magnitudes above the 50\%
completeness limit, to mimic the effect of placing M33 at the distance
of NGC~300.
We found a smaller scale length evolution than \citet{Williams2009a},
with the scale length increasing from $1.0 \pm 0.2$ kpc at 8 Gyr to
$1.3 \pm 0.1$ kpc at 0.1 Gyr.  
In contrast, \citet{Williams2009a} found an increase
of $1.0 \pm 0.1$ at 10 Gyr to $1.8 \pm 0.1$ at 0.6 Gyr with the deeper
data.
Thus, the reduced depth can lead to an underestimate of the scale
length evolution; it may be possible that the evolution of NGC~300
is comparably dramatic to that seen in M33,
but that we are not sensitive to it due to the shallower depth of the
NGC~300 data.
On the other hand, the inferred evolution for the
shallower M33 data is still somewhat larger than we observed in NGC~300.
We emphasize that the difference in mean age of the two galaxies, as
seen in \S\ref{sec:depth}, is still secure.

\begin{deluxetable*}{ccccc}
\tablewidth{0pt}
\tablecaption{\label{tbl:10915}
ANGST data: ages, SFRs, and metallicities for each radial bin}

\tablehead{
\colhead{Radius} & 
\colhead{Area} & 
\colhead{Age} & 
\colhead{SFR} &
\colhead{$\mh$}\\
\colhead{(kpc)} & 
\colhead{(kpc$^2$)} & 
\colhead{(Gyr)} & 
\colhead{(\sfr)} &
\colhead{}
}

\startdata

0--0.9 & 1.40 &
0.004--0.079 & $0.0110 \pm 0.0013$ & $-0.24 \pm 0.05$\\
 & &
0.079--0.13 & $0.0069 \pm 0.0009$ & $-0.24 \pm 0.05$\\
 & &
0.13--0.25 & $0.0054 \pm 0.0011$ & $-0.25 \pm 0.05$\\
 & &
0.25--0.40 & $0.0511 \pm 0.0109$ & $-0.23 \pm 0.05$\\
 & &
0.40--10 & $0.0219 \pm 0.0085$ & $-0.51 \pm 0.05$\\
 & &
10--14 & $0.1098 \pm 0.0291$ & $-0.92 \pm 0.10$\\
\hline
0.9--1.8 & 2.53 &
0.004--0.079 & $0.0189 \pm 0.0023$ & $-0.34 \pm 0.05$\\
 & &
0.079--0.13 & $0.0117 \pm 0.0018$ & $-0.34 \pm 0.05$\\
 & &
0.13--0.20 & $0.0100 \pm 0.0009$ & $-0.34 \pm 0.05$\\
 & &
0.20--0.32 & $0.0102 \pm 0.0012$ & $-0.35 \pm 0.05$\\
 & &
0.32--0.50 & $0.0410 \pm 0.0107$ & $-0.34 \pm 0.05$\\
 & &
0.50--10 & $0.0041 \pm 0.0036$ & $-0.47 \pm 0.06$\\
 & &
10--14 & $0.1358 \pm 0.0488$ & $-0.78 \pm 0.16$\\
\hline
1.8--2.7 & 1.98 &
0.004--0.079 & $0.0089 \pm 0.0015$ & $-0.38 \pm 0.05$\\
 & &
0.079--0.13 & $0.0069 \pm 0.0016$ & $-0.38 \pm 0.05$\\
 & &
0.13--0.20 & $0.0077 \pm 0.0008$ & $-0.38 \pm 0.05$\\
 & &
0.20--0.32 & $0.0081 \pm 0.0005$ & $-0.38 \pm 0.05$\\
 & &
0.32--0.50 & $0.0162 \pm 0.0019$ & $-0.38 \pm 0.06$\\
 & &
0.50--1.6 & $0.0064 \pm 0.0009$ & $-0.38 \pm 0.06$\\
 & &
1.6--10 & $0.0046 \pm 0.0043$ & $-0.51 \pm 0.08$\\
 & &
10--14 & $0.0452 \pm 0.0141$ & $-0.82 \pm 0.14$\\
\hline
2.7--3.6 & 1.90 &
0.004--0.079 & $0.0124 \pm 0.0024$ & $-0.47 \pm 0.11$\\
 & &
0.079--0.13 & $0.0074 \pm 0.0008$ & $-0.48 \pm 0.11$\\
 & &
0.13--0.20 & $0.0054 \pm 0.0010$ & $-0.48 \pm 0.11$\\
 & &
0.20--0.40 & $0.0045 \pm 0.0005$ & $-0.48 \pm 0.11$\\
 & &
0.40--0.63 & $0.0090 \pm 0.0012$ & $-0.48 \pm 0.11$\\
 & &
0.63--2.0 & $0.0049 \pm 0.0025$ & $-0.49 \pm 0.12$\\
 & &
2.0--10 & $0.0035 \pm 0.0029$ & $-0.60 \pm 0.06$\\
 & &
10--14 & $0.0162 \pm 0.0060$ & $-0.92 \pm 0.16$\\
\hline
3.6--4.5 & 1.83 &
0.004--0.100 & $0.0050 \pm 0.0009$ & $-0.41 \pm 0.07$\\
 & &
0.100--0.158 & $0.0034 \pm 0.0004$ & $-0.42 \pm 0.07$\\
 & &
0.158--0.25 & $0.0035 \pm 0.0004$ & $-0.42 \pm 0.07$\\
 & &
0.25--0.40 & $0.0044 \pm 0.0008$ & $-0.41 \pm 0.07$\\
 & &
0.40--0.79 & $0.0045 \pm 0.0005$ & $-0.43 \pm 0.07$\\
 & &
0.79--3.2 & $0.0022 \pm 0.0011$ & $-0.43 \pm 0.07$\\
 & &
3.2--10 & $0.0042 \pm 0.0031$ & $-0.68 \pm 0.04$\\
 & &
10--14 & $0.0052 \pm 0.0025$ & $-1.29 \pm 0.31$\\
\hline
4.5--5.4 & 1.35 &
0.004--0.079 & $0.0027 \pm 0.0006$ & $-0.58 \pm 0.10$\\
 & &
0.079--0.158 & $0.0015 \pm 0.0002$ & $-0.58 \pm 0.09$\\
 & &
0.158--0.25 & $0.0017 \pm 0.0003$ & $-0.58 \pm 0.10$\\
 & &
0.25--0.40 & $0.0014 \pm 0.0002$ & $-0.58 \pm 0.09$\\
 & &
0.40--0.79 & $0.0022 \pm 0.0005$ & $-0.59 \pm 0.10$\\
 & &
0.79--1.6 & $0.0033 \pm 0.0003$ & $-0.60 \pm 0.09$\\
 & &
1.6--5.0 & $0.0023 \pm 0.0012$ & $-0.71 \pm 0.07$\\
 & &
5.0--14 & $0.0011 \pm 0.0005$ & $-1.21 \pm 0.23$
\enddata
\end{deluxetable*}

\begin{deluxetable*}{ccccc}
\tablewidth{0pt}
\tablecaption{\label{tbl:fits}
Fits to the surface density and metallicity gradient}
\tablehead{
\colhead{Age} & \colhead{$\Sigma_0$} & \colhead{$R_d$} &
  \colhead{$Z_0$} & \colhead{$\Delta Z/ \Delta r$}\\
\colhead{(Gyr)} & \colhead{(\Msun\ pc$^{-2}$)} & \colhead{(kpc)} & &
  \colhead{(kpc$^{-1}$)}
}
\startdata
0.004 & $506.946 \pm 106.331$ & $1.21 \pm 0.10$ & $-0.23 \pm 0.05$ & $-0.072 \pm 0.019$\\
0.1   & $507.602 \pm 106.786$ & $1.21 \pm 0.10$ & $-0.23 \pm 0.05$ & $-0.075 \pm 0.019$\\
1.0   & $513.618 \pm 110.948$ & $1.16 \pm 0.10$ & $-0.40 \pm 0.05$ & $-0.057 \pm 0.018$\\
5.0   & $586.142 \pm 129.738$ & $0.95 \pm 0.09$ & $-0.52 \pm 0.05$ & $-0.078 \pm 0.020$\\
10    & $443.474 \pm 110.079$ & $0.97 \pm 0.10$ & $-0.89 \pm 0.11$ & $-0.059 \pm 0.049$
\enddata
\end{deluxetable*}

\subsection{Metallicity}
\label{sec:metallicity}

In most models of galaxy evolution, the buildup of a stellar
population is accompanied by an increase in the mean stellar
metallicity.  Outside the Milky Way, metallicity is often measured
only for the youngest stellar populations, using either \HII\ regions 
or atmospheres of A and B stars (with the latter method being used only for the nearest galaxies).  Here
we discuss the present-day metallicity structure of NGC~300 and
compare it to its past evolution as derived from stellar populations.

\subsubsection{Present Day Metallicity}

A radial metallicity gradient for NGC~300 in the gas and young stars has been reported by a number of
authors
\citep{Pagel1979,Webster1983,Edmunds1984,Deharveng1988,Zaritsky1994,Urbaneja2005,Kudritzki2008,Bresolin2009}.
All these studies found that metallicity is highest in the
center and decreases toward larger
radii, although the overall metallicity depends on the
calibration method used.  The results best suited for comparison with
metallicities derived from our stellar populations are those of
\citet{Urbaneja2005} and \citet{Kudritzki2008}, since they determine
metallicities using stellar spectroscopy of individual
young A and B stars from the galactic center out to $\sim6.8$ kpc, and
are thus measuring the stellar, rather than gas-phase, metallicity gradient.  
The best-fit abundance gradient of \citet{Kudritzki2008} is
\begin{equation}
\label{eqn:metal}
[Z] = (-0.06 \pm 0.09) - (0.078 \pm 0.021)r
\end{equation}
where $r$ is radius in kpc (assuming the distance used in this paper of
2.0 Mpc) and [Z] is an average metallicity based on a combination of
multiple abundance measurements (mostly Fe, Ti, and Cr lines).
We adopt Equation~\ref{eqn:metal} as the best estimate of the present-day metallicity gradient.
The \HII-region metallicities from
temperature-sensitive emission lines presented in \citet{Bresolin2009}
also agree remarkably well with these stellar metallicities.

\subsubsection{Metallicity Evolution from HST Data}

\begin{figure}
\plotone{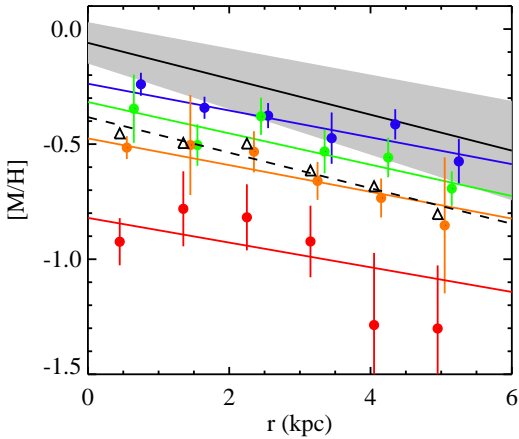}
\caption{\label{fig:zgradient}
  Solid black line: best-fit metallicity gradient obtained from young
  massive stars
  \citep{Kudritzki2008}, with shaded region showing uncertainties.
  Blue circles are metallicities measured from our CMDs in each radial bin at the present
  day (i.e., the mean metallicity for the past 100 Myr);  green,
  orange, and red circles are metallicities for the 1--5 Gyr, 5--10 Gyr, and 10--14
  Gyr bins, respectively.  
  Solid lines of corresponding colors show the best linear fit for
  each set of points.
  Errors are calculated as in Figure~\ref{fig:sfh}.
  Mean metallicities for the entire stellar
  population are shown with black triangles, with a dashed black line
  showing the best linear fit.
  Points have been offset in radius to avoid overlapping error bars.}
\end{figure}

The metallicity gradient measured in young stars by \citet{Kudritzki2008}
can be compared to the present-day metallicities inferred from the
observed stellar populations.
The CMDs contain information on the metallicities as well as the ages
of stars, especially for stars off the main sequence
\citep{Gallart2005}.
A full suite of isochrones with varying age
\emph{and} metallicity is used to fit each CMD, so that in the final
SFH each time bin is associated with a particular metallicity.
The full metallicity history for all radial bins is shown in
Figure~\ref{fig:metal}.  
Uncertainties in the metallicity may result from several factors: 1)
an age-metallicity degeneracy in the RGB, 2) photometric errors in the
CMD, 3) uncertainties in the reddening,
4) uncertainties in modeling stars on the instability strip, and 5)
errors in the isochrones, especially for supergiants and AGB stars.
We address these uncertainties by constraining the metallicity not to
decrease significantly with time, which eliminates unphysical solutions to the SFH.

We tested the effect of this parameter by deriving the SFH without it
as well.  Not constraining the
metallicity results in lower \mh\ values for intermediate age stars
(1--10 Gyr), especially in the most crowded regions, but the effect on
the SFR is minimal.  The trends in SFH reported in this paper do not
depend on whether or not the metallicity constraint is enforced.

The derived SFH from the ANGST data reproduces a metallicity gradient
in the present-day metallicity (4--100 Myr) time bin.
In Figure~\ref{fig:zgradient} we show
the \citet{Kudritzki2008} metallicity gradient along with our
derived metallicity gradient.
Fit parameters for the linear fits are given in Table~\ref{tbl:fits}.
The MATCH-derived present-day metallicities are generally consistent with
the observed values within the error bars, although the metallicities are all
systematically low and fall near the minimum value in the observed uncertainty range.  The mean
metallicities of the entire population are lower as well, as would be expected for a stellar
population that built up gradually over $\sim10$ Gyr.  However, the
mean stellar metallicity still shows a radial gradient at all epochs.

Since the color of the main sequence is not significantly affected by
metallicity, our CMD-fitting code may not be
sensitive to a metallicity increase at very recent times.
The metallicities we find are
therefore a lower limit on the current values (at least in the central
regions where the infall of unenriched gas is assumed to have ended).
This insensitivity is especially true in the innermost
radial bins, in which the low current SFRs and the effects of crowding
make the present-day metallicity especially difficult to determine.  Not
coincidentally, these values fall more significantly below the
\citet{Kudritzki2008} results than do the values for the outer disk.

The change in the metallicity gradient over time can be seen in the mean
metallicity derived during multiple time bins, also shown in
Figure~\ref{fig:zgradient}.  Stellar population ages of 4--100 Myr,
1--5 Gyr, 5--10 Gyr, and 10--14 Gyr are shown.  (The metallicity values for
age 100 Myr -- 1 Gyr are essentially identical to the present-day
values.)
Results of linear fits to the gradients are presented in Table~\ref{tbl:fits}.
Outside 2 kpc, the slope of the metallicity
gradient has remained fairly constant over time, with the overall
metallicity increasing.  
In the central 2 kpc, we see a significant
metallicity increase from the 10--14 Gyr bin to the 5--10 Gyr bin, which
corresponds to the epoch when significant star formation activity was
taking place in the galactic center.

\subsubsection{Comparison with M33's metallicity}

Figure~\ref{fig:zgradient} shows that NGC~300 has both a significant
radial metallicity gradient and a significant increase in metallicity
with time.  
Although we constrained the metallicity not to decrease with time,
there was no barrier to its remaining constant in deriving the SFHs,
or to its taking arbitrary slopes with radius.
In contrast, M33 has a shallower gradient and metallicity
which has remained nearly constant with time \citep[][Holtzman et
  al. 2009, in prep.]{Magrini2009}.

These results are consistent with chemical evolution models.  Ongoing
star formation in M33 may indicate gas infall, which can dilute the
metallicity, keeping it constant with time despite the continuing
enrichment from star formation.  
In contrast, NGC~300 formed a small
fraction of its stars at late times, which may indicate a less
important role for ongoing gas accretion, and thus chemical evolution
that behaves more like a closed box.  This may also be a
counterexample to the ``downsizing'' model, since NGC~300 is a fairly
low-mass disk galaxy.

\section{Conclusions}
\label{sec:conclusions}

We have presented resolved stellar photometry of 3 \emph{HST}/ACS
fields in a continuous radial strip from the center of NGC~300 to 5.4
kpc, as well as additional archival fields scattered throughout the
disk.  For both the ANGST (continuous) and archival fields, we have
divided the stars into 6 radial bins of width 0.9 kpc and used
CMD-fitting to derive the SFH in each bin.
We can resolve the red clump in the majority of the radial bins.
We find that the disk of NGC~300 has a dominant old population
throughout the observed region: looking at the disk as a whole,
$\sim$80\% of stars are older than 6 Gyr.
However, the outer
parts of the disk have a higher percentage of young stars than the
inner parts, consistent with inside-out growth.
In the inner regions,
$>90$\% of the stars are older than 6 Gyr, while in the outermost
radial bin ($4.5 < 5.4$ kpc), only $\sim$40\% of stars are this old.
Comparison of formation and final positions of stars in an $N$-body
simulation of an NGC~300-like galaxy indicates that the effects of
migration should not be substantial in a galaxy of this size.

We calculated the surface density profile of the disk at each time step by
summing over the stellar mass formed and fitting the profile to derive
the corresponding scale
lengths.  According to inside-out growth, the scale length of the disk
should increase with time.  While we found this to be the case, the
increase is relatively small, from $\sim1.1$ to $\sim1.3$ kpc over the lifetime of
the disk.  In contrast, the
scale lengths of M33 found with an identical method show a much
greater increase, from 1.0 kpc to 1.8 kpc \citep{Williams2009a}, more
in line with the theoretical predictions of \citet{Mo1998}. 

Although M33 is nearly a twin to NGC~300 in Hubble type and mass,
there are differences between the two galaxies which may be related to
their different SFHs.
The primary difference in morphology is the
presence of a disk break in M33.  
Breaks have been shown to be extremely
common in disk galaxies \citep{Pohlen2006}, and M33 is no exception,
with a break at 8 kpc \citep{Ferguson2007}.  The scale length increase seen by
\citet{Williams2009a} only holds inside the disk break; outside the
break the opposite trend is seen, with a decreasing scale length with
time.  In contrast, NGC~300 is unusual in that it has no disk break out
to at least 14 kpc \citep{Bland-Hawthorn2005}.  
\citet{Sanchez-Blazquez2009} suggest that pure exponential profiles
may be a feature of galaxies that have undergone less radial
mixing, which is consistent with our analysis in
\S\ref{sec:migration} that NGC~300 has undergone relatively little
recent migration.

Environmental factors may also be important, as NGC~300 is isolated
from other large galaxies while M33 appears to be interacting with M31
\citep{Braun2004,Bekki2008,McConnachie2009,Putman2009}.
An influx of gas onto M33 in recent times may have triggered star
formation in the outskirts of M33 and contributed to the growth of the
disk. 

Finally, we note that despite the unbroken exponential profile of
NGC~300, it is not without features in the very outer disk (beyond the
extent of our observations): \citet{Vlajic2009} find a change in the
abundance gradient at $\sim10$ kpc.  They  propose two possible
explanations for the upturn in the gradient: radial mixing, which we
conclude is unlikely to be the sole effect based on the simulation presented in
\S\ref{sec:migration}, and an accretion scenario in which the outer
regions of the disk form stars later and become enriched, flattening
the gradient.
We suggest
that additional \emph{HST} observations of resolved stars farther out
in NGC~300 would provide useful constraints on the total disk scale
length and evolution.

From our derived SFH for NGC~300, we find a present-day metallicity
gradient roughly consistent with observational results
\citep{Kudritzki2008,Bresolin2009}.  Additionally, we find that the
metallicity has increased with time in all radial bins, suggesting
a lack of infall of unenriched gas.  This is broadly consistent with
our finding that the majority of stars in NGC~300 formed prior to 6
Gyr ago, as more recent gas infall would likely have triggered more
enhanced recent star formation.  As M33 has been shown to have little
evolution in metallicity and more of its stars formed recently, it may
have experienced more gas infall at later times.  Despite the visual
similarities between these two galaxies, they seem to have markedly
different histories.

\acknowledgements
We thank Dennis Zaritsky for helpful comments.
We also thank the anonymous referee for several useful suggestions.
Support for this work was provided by NASA through grant GO-10915
from the Space Telescopes Science Institute, which is operated by the
Association of Universities for Research in Astronomy, Inc., under
NASA contract NAS5-26555.
J.J.D.\ was partially supported as a Wycoff Fellow.
This research has made use of the NASA/IPAC Extragalactic Database
(NED) which is operated by JPL/Caltech, under contract with NASA.
The Digitized Sky Surveys were produced at the Space Telescope Science
Institute under U.S. Government grant NAG W-2166. The images of these
surveys are based on photographic data obtained using the Oschin
Schmidt Telescope on Palomar Mountain and the UK Schmidt
Telescope. The plates were processed into the present compressed
digital form with the permission of these institutions.
The UK Schmidt Telescope was operated by the Royal Observatory
Edinburgh, with funding from the UK Science and Engineering Research
Council (later the UK Particle Physics and Astronomy Research
Council), until 1988 June, and thereafter by the Anglo-Australian
Observatory. 

{\it Facilities:} \facility{HST (ACS)}

\appendix
\section{Archival Data}
\label{sec:archival}

\begin{figure*}
\plotone{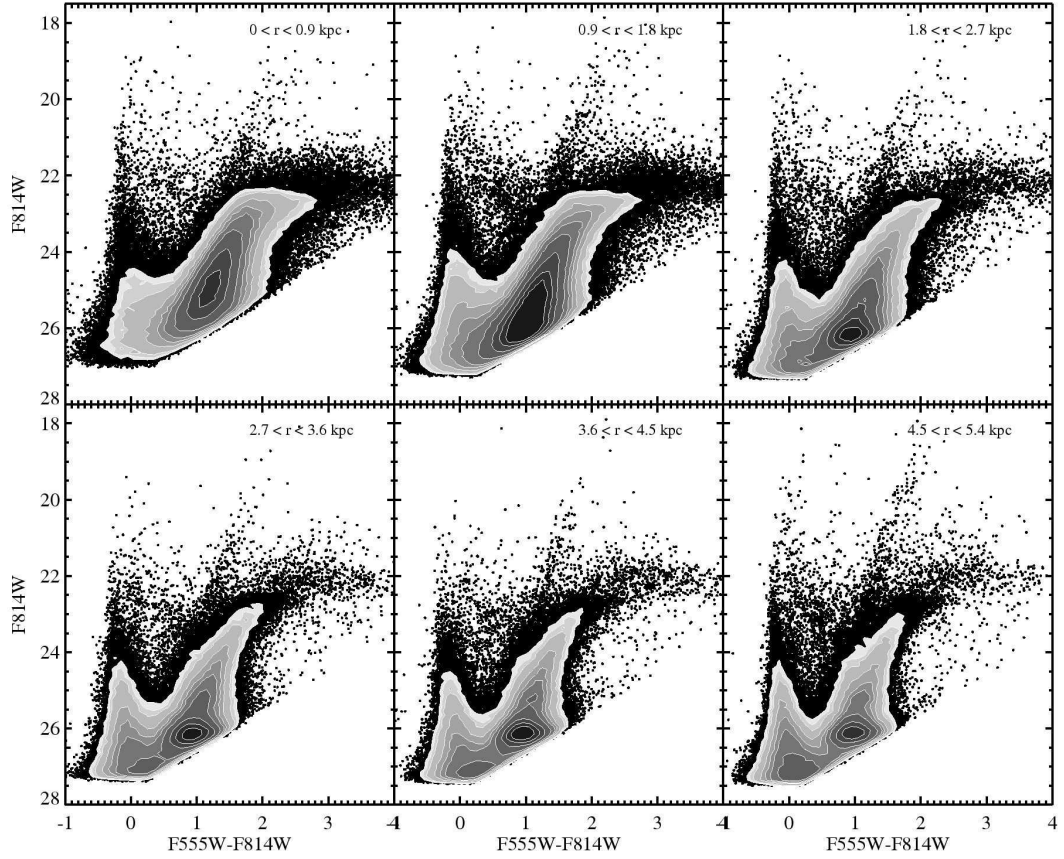}
\caption{\label{fig:cmds_9492}
  CMDs of radial bins from archival data.  Quality cuts are identical
  to those described in Figure~\ref{fig:cmds}.}
\end{figure*}

\begin{figure*}
\plotone{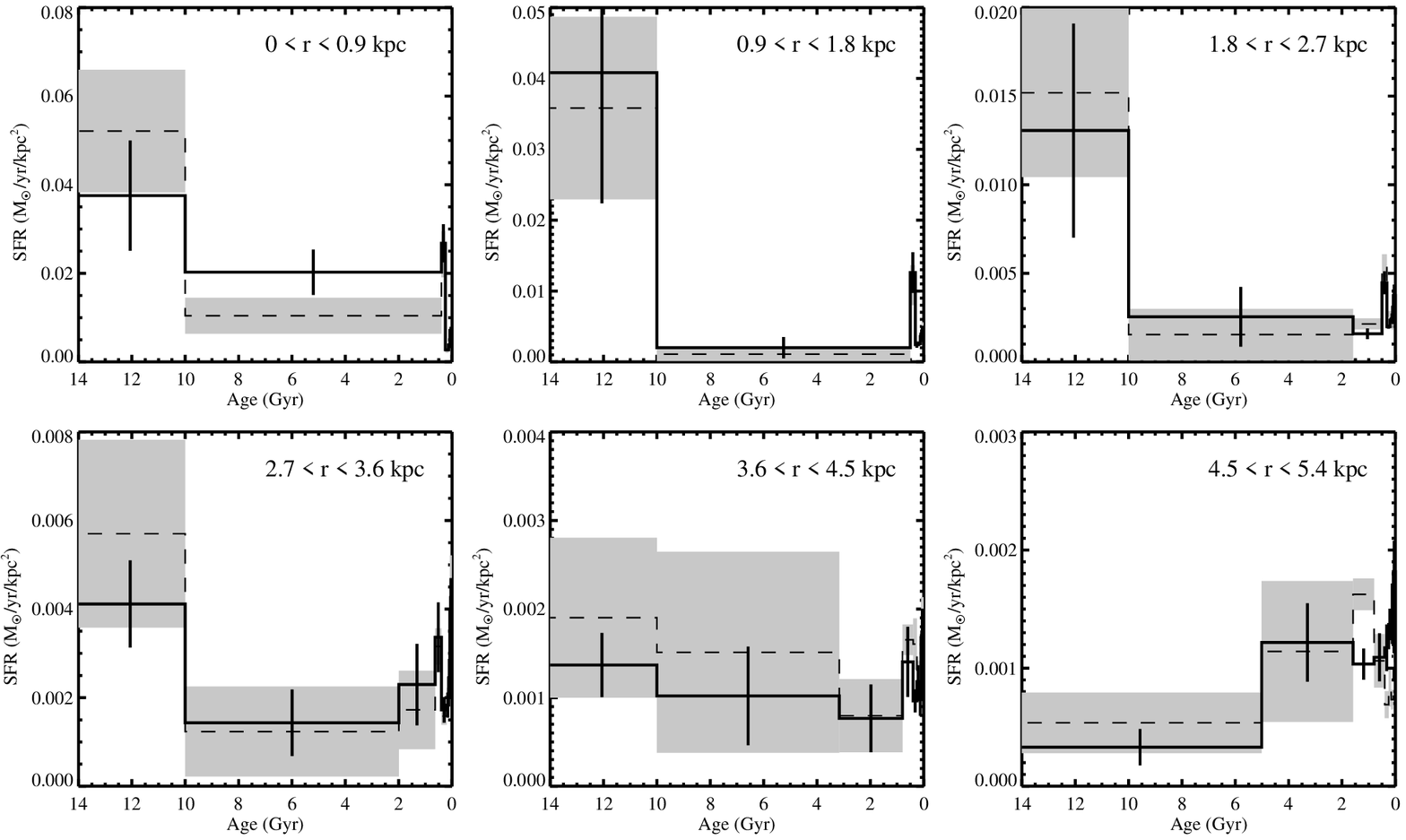}
\caption{\label{fig:sfh_9492}
  Lifetime SFHs of radial bins from archival data (solid lines).  
  Errors are calculated as in Figure~\ref{fig:sfh}.
  For comparison, the ANGST SFHs are shown as dashed lines, with
  uncertainties shown as shaded regions.}
\end{figure*}

\begin{figure*}
\plotone{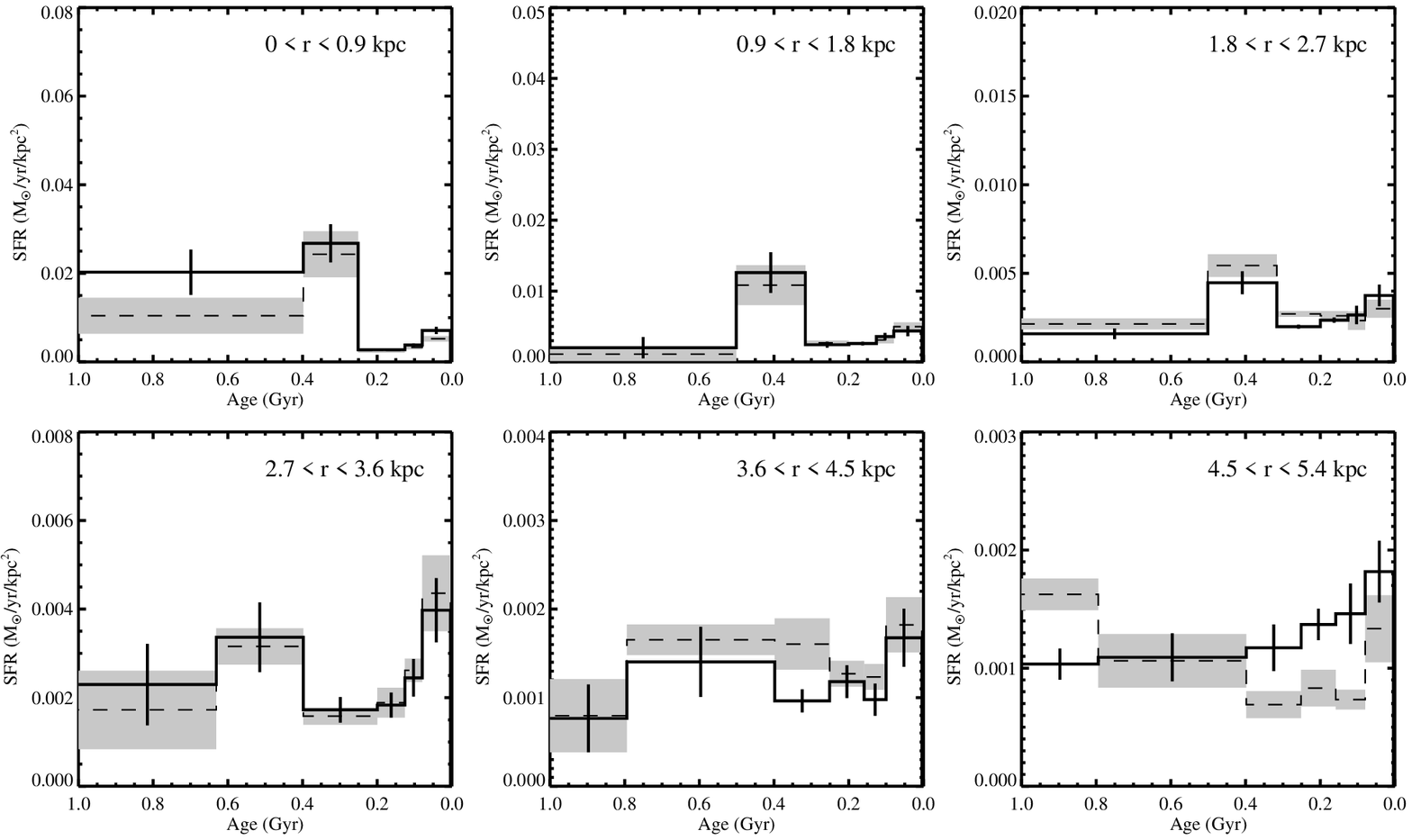}
\caption{\label{fig:sfh_recent_9492}
  Recent ($<1$ Gyr) SFHs of radial bins from archival data (solid lines).
  Errors are calculated as in Figure~\ref{fig:sfh}.
  For comparison, the ANGST SFHs are shown as dashed lines, with
  uncertainties shown as shaded regions.}
\end{figure*}
 
\begin{figure}
\epsscale{0.5}
\plotone{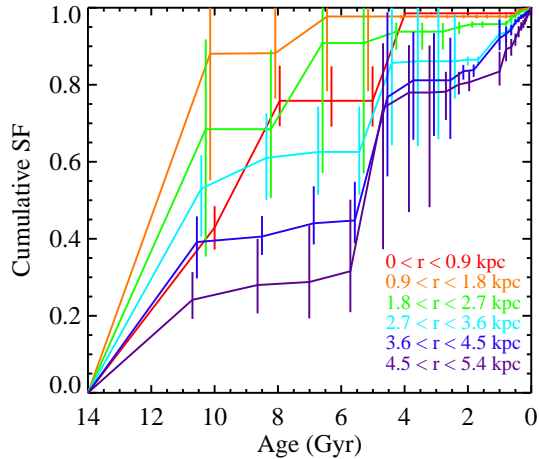}
\caption{\label{fig:cumul_9492}
  Cumulative star formation for all radial bins for archival data.
  Bin edges have been offset to avoid overlapping error bars.}
\end{figure}

\begin{figure}
\epsscale{0.5}
\plotone{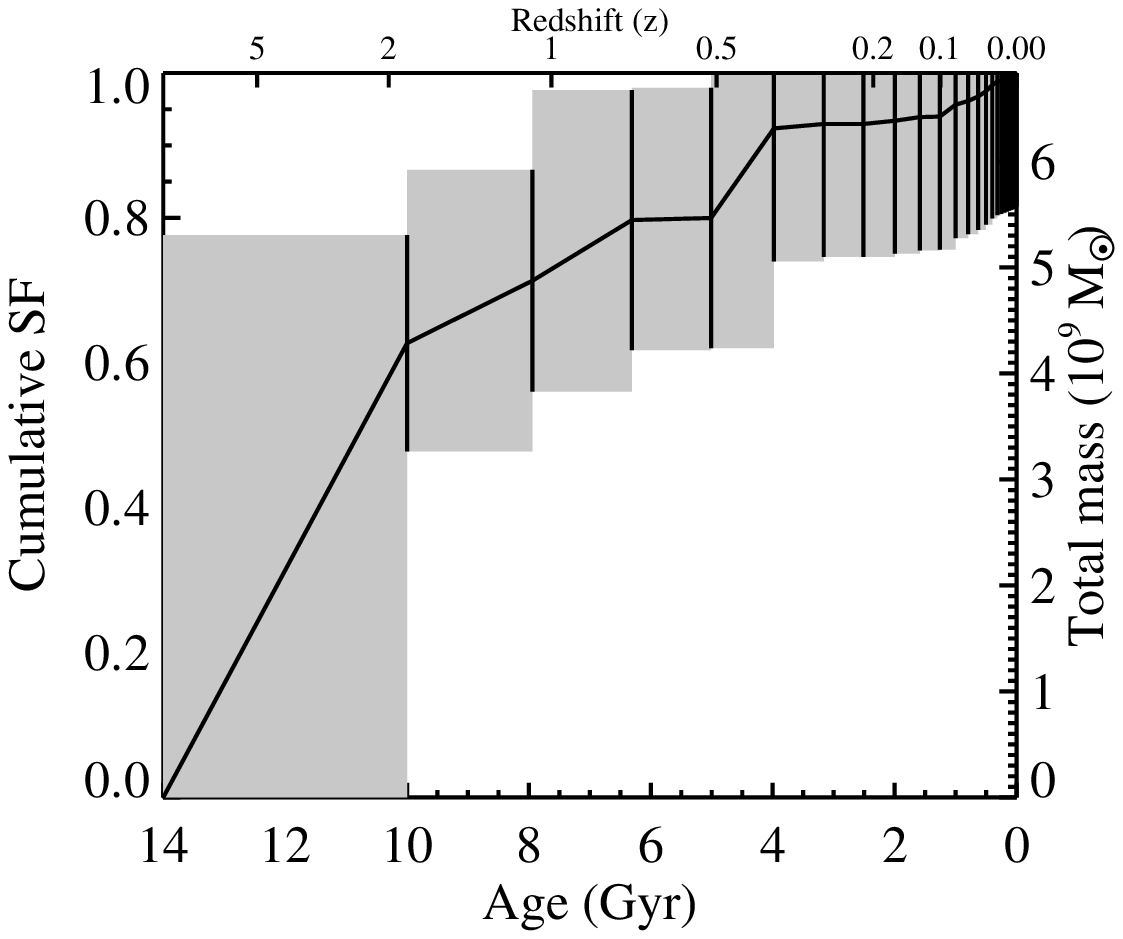}
\caption{\label{fig:cumul_global_9492}
  Cumulative star formation for the entire galaxy out to 5.4 kpc.
  Shaded region indicates uncertainty.}
\end{figure}

\begin{figure*}
\epsscale{1.0}
\plotone{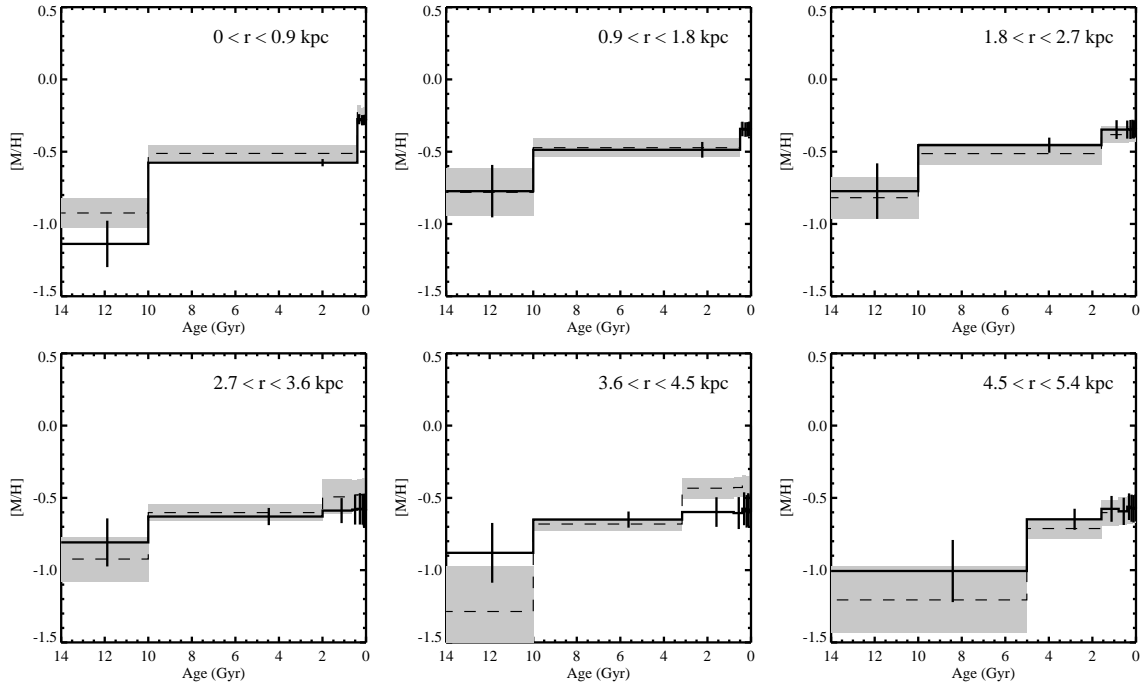}
\caption{\label{fig:metal_9492}
  Mean metallicities corresponding to SFHs in Figure~\ref{fig:sfh_9492}.
  Solid lines are archival data, and dashed lines (with shaded uncertainties) are ANGST data.
  Errors are calculated as in Figure~\ref{fig:sfh}.}
\end{figure*}

\begin{deluxetable*}{ccccc}
\tablewidth{0pt}
\tablecaption{\label{tbl:9492}
Archival data: ages, SFRs, and metallicities for each radial bin}

\tablehead{
\colhead{Radius} & 
\colhead{Area} & 
\colhead{Age} & 
\colhead{SFR} &
\colhead{$\mh$}\\
\colhead{(kpc)} & 
\colhead{(kpc$^2$)} & 
\colhead{(Gyr)} & 
\colhead{(\sfr)} &
\colhead{}
}

\startdata
0--0.9 & 1.89 &
0.004--0.079 & $0.0201 \pm 0.0023$ & $-0.28 \pm 0.03$\\
 & &
0.079--0.13 & $0.0108 \pm 0.0010$ & $-0.28 \pm 0.04$\\
 & &
0.13--0.25 & $0.0078 \pm 0.0006$ & $-0.28 \pm 0.04$\\
 & &
0.25--0.40 & $0.0760 \pm 0.0122$ & $-0.28 \pm 0.03$\\
 & &
0.40--10 & $0.0575 \pm 0.0145$ & $-0.58 \pm 0.03$\\
 & &
10--14 & $0.1065 \pm 0.0352$ & $-1.14 \pm 0.16$\\
\hline	    
0.9--1.8 & 3.41 &
0.004--0.079 & $0.0223 \pm 0.0037$ & $-0.34 \pm 0.05$\\
 & &
0.079--0.13 & $0.0183 \pm 0.0026$ & $-0.35 \pm 0.05$\\
 & &
0.13--0.20 & $0.0133 \pm 0.0013$ & $-0.35 \pm 0.05$\\
 & &
0.20--0.32 & $0.0124 \pm 0.0021$ & $-0.35 \pm 0.05$\\
 & &
0.32--0.50 & $0.0645 \pm 0.0146$ & $-0.34 \pm 0.05$\\
 & &
0.50--10 & $0.0102 \pm 0.0076$ & $-0.49 \pm 0.05$\\
 & &
10--14 & $0.2088 \pm 0.0944$ & $-0.77 \pm 0.18$\\
\hline	     
1.8--2.7 & 2.47 &
0.004--0.079 & $0.0139 \pm 0.0023$ & $-0.34 \pm 0.06$\\
 & &
0.079--0.13 & $0.0098 \pm 0.0019$ & $-0.35 \pm 0.06$\\
 & &
0.13--0.20 & $0.0087 \pm 0.0006$ & $-0.34 \pm 0.06$\\
 & &
0.20--0.32 & $0.0074 \pm 0.0004$ & $-0.35 \pm 0.07$\\
 & &
0.32--0.50 & $0.0165 \pm 0.0024$ & $-0.35 \pm 0.06$\\
 & &
0.50--1.6 & $0.0059 \pm 0.0011$ & $-0.35 \pm 0.06$\\
 & &
1.6--10 & $0.0094 \pm 0.0062$ & $-0.46 \pm 0.05$\\
 & &
10--14 & $0.0483 \pm 0.0223$ & $-0.77 \pm 0.19$\\
\hline	     
2.7--3.6 & 2.71 &
0.004--0.079 & $0.0162 \pm 0.0029$ & $-0.58 \pm 0.11$\\
 & &
0.079--0.13 & $0.0100 \pm 0.0017$ & $-0.59 \pm 0.11$\\
 & &
0.13--0.20 & $0.0074 \pm 0.0011$ & $-0.58 \pm 0.11$\\
 & &
0.20--0.40 & $0.0070 \pm 0.0012$ & $-0.58 \pm 0.11$\\
 & &
0.40--0.63 & $0.0137 \pm 0.0032$ & $-0.58 \pm 0.10$\\
 & &
0.63--2.0 & $0.0093 \pm 0.0037$ & $-0.59 \pm 0.09$\\
 & &
2.0--10 & $0.0058 \pm 0.0031$ & $-0.63 \pm 0.06$\\
 & &
10--14 & $0.0167 \pm 0.0040$ & $-0.81 \pm 0.17$\\
\hline	     
3.6--4.5 & 3.94 &
0.004--0.100 & $0.0099 \pm 0.0019$ & $-0.59 \pm 0.11$\\
 & &
0.100--0.16 & $0.0058 \pm 0.0011$ & $-0.58 \pm 0.11$\\
 & &
0.16--0.25 & $0.0070 \pm 0.0011$ & $-0.59 \pm 0.11$\\
 & &
0.25--0.40 & $0.0057 \pm 0.0008$ & $-0.58 \pm 0.11$\\
 & &
0.40--0.79 & $0.0083 \pm 0.0023$ & $-0.61 \pm 0.11$\\
 & &
0.79--3.2 & $0.0045 \pm 0.0023$ & $-0.60 \pm 0.10$\\
 & &
3.2--10 & $0.0060 \pm 0.0033$ & $-0.65 \pm 0.06$\\
 & &
10--14 & $0.0081 \pm 0.0021$ & $-0.88 \pm 0.21$\\
\hline	     
4.5--5.4 & 4.68 &
0.004--0.079 & $0.0128 \pm 0.0018$ & $-0.57 \pm 0.09$\\
 & &
0.079--0.16 & $0.0103 \pm 0.0018$ & $-0.58 \pm 0.09$\\
 & &
0.16--0.25 & $0.0096 \pm 0.0009$ & $-0.57 \pm 0.09$\\
 & &
0.25--0.40 & $0.0082 \pm 0.0014$ & $-0.56 \pm 0.09$\\
 & &
0.40--0.79 & $0.0077 \pm 0.0014$ & $-0.59 \pm 0.09$\\
 & &
0.79--1.6 & $0.0073 \pm 0.0009$ & $-0.58 \pm 0.09$\\
 & &
1.6--5.0 & $0.0085 \pm 0.0023$ & $-0.65 \pm 0.07$\\
 & &
5.0--14 & $0.0023 \pm 0.0011$ & $-1.01 \pm 0.21$
\enddata
\end{deluxetable*}

In this section we present the results of analyzing archival fields in
NGC~300 (see \S\ref{sec:imaging}) in the same manner as the ANGST data.
CMDs for the archival data are shown in Figure~\ref{fig:cmds_9492}.
The quality cuts used for selecting stars were identical to those
described for the ANGST data in \S\ref{sec:photometry}.
Although the archival fields are not radially aligned, the stars in
these fields can be sorted into the same 6 bins as the new
observations.  
These fields are not evenly spaced
in radius, as shown in Figure~\ref{fig:overlays}, and were placed on
regions of high current star formation.
We include only stars up to a galactocentric distance of 5.4 kpc, the
radial extent of the ANGST data.

As with the ANGST data, we fixed the distance for these SFH
derivations at $m-M=26.43$.
The results of the fits were combined into time bins identical to
those used for the ANGST data.
For the archival data, mean extinction values for each bin are, from
inner to outer:
$A_V = 0.30 \pm 0.05, 0.20 \pm 0.05, 0.10 \pm 0.05, 0.10 \pm
0.05, 0.10 \pm 0.05, 0.13 \pm 0.07$.  The higher extinction values for
the archival data are likely due to the location of these fields in
regions of higher star formation, which should correspond with
increased dust content that affects older stars as well.  
Indeed, \citet{Roussel2005} studied extinction in NGC~300 and found
that extinction is variable for young clusters.

Reassuringly, the SFH from the archival data, shown in Figures
\ref{fig:sfh_9492} and \ref{fig:sfh_recent_9492},
is very similar to that
derived from the ANGST data, despite the different filters and
depths.
Cumulative SFHs for archival data divided into radial bins are shown
in Figure~\ref{fig:cumul_9492}, and integrated for the whole galaxy in
Figure~\ref{fig:cumul_global_9492}.  Metallicities are shown in
Figure~\ref{fig:metal_9492}.
Table~\ref{tbl:9492} gives the full SFH for the archival data.


\begin{thebibliography}{77}
\expandafter\ifx\csname natexlab\endcsname\relax\def\natexlab#1{#1}\fi

\bibitem[{{Azzollini} {et~al.}(2008){Azzollini}, {Trujillo}, \&
  {Beckman}}]{Azzollini2008}
{Azzollini}, R., {Trujillo}, I., \& {Beckman}, J.~E. 2008, \apjl, 679, L69

\bibitem[{{Barden} {et~al.}(2005){Barden}, {Rix}, {Somerville}, {Bell},
  {H{\"a}u{\ss}ler}, {Peng}, {Borch}, {Beckwith}, {Caldwell}, {Heymans},
  {Jahnke}, {Jogee}, {McIntosh}, {Meisenheimer}, {S{\'a}nchez}, {Wisotzki}, \&
  {Wolf}}]{Barden2005}
{Barden}, M. {et~al.} 2005, \apj, 635, 959

\bibitem[{{Bekki}(2008)}]{Bekki2008}
{Bekki}, K. 2008, \mnras, 390, L24

\bibitem[{{Bertelli} {et~al.}(1994){Bertelli}, {Bressan}, {Chiosi}, {Fagotto},
  \& {Nasi}}]{Bertelli1994}
{Bertelli}, G., {Bressan}, A., {Chiosi}, C., {Fagotto}, F., \& {Nasi}, E. 1994,
  \aaps, 106, 275

\bibitem[{{Bland-Hawthorn} {et~al.}(2005){Bland-Hawthorn}, {Vlaji{\'c}},
  {Freeman}, \& {Draine}}]{Bland-Hawthorn2005}
{Bland-Hawthorn}, J., {Vlaji{\'c}}, M., {Freeman}, K.~C., \& {Draine}, B.~T.
  2005, \apj, 629, 239

\bibitem[{{Boissier} \& {Prantzos}(1999)}]{Boissier1999}
{Boissier}, S., \& {Prantzos}, N. 1999, \mnras, 307, 857

\bibitem[{{Braun} \& {Thilker}(2004)}]{Braun2004}
{Braun}, R., \& {Thilker}, D.~A. 2004, \aap, 417, 421

\bibitem[{{Bresolin} {et~al.}(2009){Bresolin}, {Gieren}, {Kudritzki},
  {Pietrzy{\'n}ski}, {Urbaneja}, \& {Carraro}}]{Bresolin2009}
{Bresolin}, F., {Gieren}, W., {Kudritzki}, R.-P., {Pietrzy{\'n}ski}, G.,
  {Urbaneja}, M.~A., \& {Carraro}, G. 2009, \apj, 700, 309

\bibitem[{{Bresolin} {et~al.}(2005){Bresolin}, {Pietrzy{\'n}ski}, {Gieren}, \&
  {Kudritzki}}]{Bresolin2005}
{Bresolin}, F., {Pietrzy{\'n}ski}, G., {Gieren}, W., \& {Kudritzki}, R.-P.
  2005, \apj, 634, 1020

\bibitem[{{Brook} {et~al.}(2006){Brook}, {Kawata}, {Martel}, {Gibson}, \&
  {Bailin}}]{Brook2006}
{Brook}, C.~B., {Kawata}, D., {Martel}, H., {Gibson}, B.~K., \& {Bailin}, J.
  2006, \apj, 639, 126

\bibitem[{{Burkert} {et~al.}(1992){Burkert}, {Truran}, \&
  {Hensler}}]{Burkert1992}
{Burkert}, A., {Truran}, J.~W., \& {Hensler}, G. 1992, \apj, 391, 651

\bibitem[{{Butler} {et~al.}(2004){Butler}, {Mart{\'{\i}}nez-Delgado}, \&
  {Brandner}}]{Butler2004}
{Butler}, D.~J., {Mart{\'{\i}}nez-Delgado}, D., \& {Brandner}, W. 2004, \aj,
  127, 1472

\bibitem[{{Carignan}(1985)}]{Carignan1985}
{Carignan}, C. 1985, \apjs, 58, 107

\bibitem[{{Chiappini} {et~al.}(1997){Chiappini}, {Matteucci}, \&
  {Gratton}}]{Chiappini1997}
{Chiappini}, C., {Matteucci}, F., \& {Gratton}, R. 1997, \apj, 477, 765

\bibitem[{{Corbelli} \& {Salucci}(2000)}]{Corbelli2000}
{Corbelli}, E., \& {Salucci}, P. 2000, \mnras, 311, 441

\bibitem[{{Dalcanton} {et~al.}(2009){Dalcanton}, {Williams}, {Seth}, {Dolphin},
  {Holtzman}, {Rosema}, {Skillman}, {Cole}, {Girardi}, {Gogarten},
  {Karachentsev}, {Olsen}, {Weisz}, {Christensen}, {Freeman}, {Gilbert},
  {Gallart}, {Harris}, {Hodge}, {de Jong}, {Karachentseva}, {Mateo}, {Stetson},
  {Tavarez}, {Zaritsky}, {Governato}, \& {Quinn}}]{Dalcanton2009}
{Dalcanton}, J.~J. {et~al.} 2009, \apjs, 183, 67

\bibitem[{{Deharveng} {et~al.}(1988){Deharveng}, {Caplan}, {Lequeux},
  {Azzopardi}, {Breysacher}, {Tarenghi}, \& {Westerlund}}]{Deharveng1988}
{Deharveng}, L., {Caplan}, J., {Lequeux}, J., {Azzopardi}, M., {Breysacher},
  J., {Tarenghi}, M., \& {Westerlund}, B. 1988, \aaps, 73, 407

\bibitem[{{Dolphin}(2000)}]{Dolphin2000}
{Dolphin}, A.~E. 2000, \pasp, 112, 1383

\bibitem[{{Dolphin}(2002)}]{Dolphin2002}
---. 2002, \mnras, 332, 91

\bibitem[{{Edmunds} \& {Pagel}(1984)}]{Edmunds1984}
{Edmunds}, M.~G., \& {Pagel}, B.~E.~J. 1984, \mnras, 211, 507

\bibitem[{{Ferguson} {et~al.}(2007){Ferguson}, {Irwin}, {Chapman}, {Ibata},
  {Lewis}, \& {Tanvir}}]{Ferguson2007}
{Ferguson}, A., {Irwin}, M., {Chapman}, S., {Ibata}, R., {Lewis}, G., \&
  {Tanvir}, N. 2007, Resolving the Stellar Outskirts of M31 and M33 (Island
  Universes - Structure and Evolution of Disk Galaxies), 239

\bibitem[{{Ford} {et~al.}(1998){Ford}, {Bartko}, {Bely}, {Broadhurst},
  {Burrows}, {Cheng}, {Clampin}, {Crocker}, {Feldman}, {Golimowski}, {Hartig},
  {Illingworth}, {Kimble}, {Lesser}, {Miley}, {Neff}, {Postman}, {Sparks},
  {Tsvetanov}, {White}, {Sullivan}, {Krebs}, {Leviton}, {La Jeunesse},
  {Burmester}, {Fike}, {Johnson}, {Slusher}, {Volmer}, \&
  {Woodruff}}]{Ford1998}
{Ford}, H.~C. {et~al.} 1998, in Society of Photo-Optical Instrumentation
  Engineers (SPIE) Conference Series, Vol. 3356, Society of Photo-Optical
  Instrumentation Engineers (SPIE) Conference Series, ed. P.~Y.~B. . J.~B.
  Breckinridge, 234--248

\bibitem[{{Gallart} {et~al.}(1999){Gallart}, {Freedman}, {Aparicio},
  {Bertelli}, \& {Chiosi}}]{Gallart1999}
{Gallart}, C., {Freedman}, W.~L., {Aparicio}, A., {Bertelli}, G., \& {Chiosi},
  C. 1999, \aj, 118, 2245

\bibitem[{{Gallart} {et~al.}(2005){Gallart}, {Zoccali}, \&
  {Aparicio}}]{Gallart2005}
{Gallart}, C., {Zoccali}, M., \& {Aparicio}, A. 2005, \araa, 43, 387

\bibitem[{{Gieren} {et~al.}(2005){Gieren}, {Pietrzy{\'n}ski}, {Soszy{\'n}ski},
  {Bresolin}, {Kudritzki}, {Minniti}, \& {Storm}}]{Gieren2005}
{Gieren}, W., {Pietrzy{\'n}ski}, G., {Soszy{\'n}ski}, I., {Bresolin}, F.,
  {Kudritzki}, R.-P., {Minniti}, D., \& {Storm}, J. 2005, \apj, 628, 695

\bibitem[{{Girardi} {et~al.}(2002){Girardi}, {Bertelli}, {Bressan}, {Chiosi},
  {Groenewegen}, {Marigo}, {Salasnich}, \& {Weiss}}]{Girardi2002}
{Girardi}, L., {Bertelli}, G., {Bressan}, A., {Chiosi}, C., {Groenewegen},
  M.~A.~T., {Marigo}, P., {Salasnich}, B., \& {Weiss}, A. 2002, \aap, 391, 195

\bibitem[{{Girardi} {et~al.}(2008){Girardi}, {Dalcanton}, {Williams}, {de
  Jong}, {Gallart}, {Monelli}, {Groenewegen}, {Holtzman}, {Olsen}, {Seth},
  {Weisz}, \& {the ANGST/ANGRRR Collaboration}}]{Girardi2008}
{Girardi}, L. {et~al.} 2008, \pasp, 120, 583

\bibitem[{{Harris} \& {Zaritsky}(2004)}]{Harris2004}
{Harris}, J., \& {Zaritsky}, D. 2004, \aj, 127, 1531

\bibitem[{{Hernandez} \& {Cervantes-Sodi}(2006)}]{Hernandez2006}
{Hernandez}, X., \& {Cervantes-Sodi}, B. 2006, \mnras, 368, 351

\bibitem[{{Hernandez} {et~al.}(1999){Hernandez}, {Valls-Gabaud}, \&
  {Gilmore}}]{Hernandez1999}
{Hernandez}, X., {Valls-Gabaud}, D., \& {Gilmore}, G. 1999, \mnras, 304, 705

\bibitem[{{Holtzman} {et~al.}(1999){Holtzman}, {Gallagher}, {Cole}, {Mould},
  {Grillmair}, {Ballester}, {Burrows}, {Clarke}, {Crisp}, {Evans}, {Griffiths},
  {Hester}, {Hoessel}, {Scowen}, {Stapelfeldt}, {Trauger}, \&
  {Watson}}]{Holtzman1999}
{Holtzman}, J.~A. {et~al.} 1999, \aj, 118, 2262

\bibitem[{{Karachentsev} {et~al.}(2003){Karachentsev}, {Grebel}, {Sharina},
  {Dolphin}, {Geisler}, {Guhathakurta}, {Hodge}, {Karachentseva}, {Sarajedini},
  \& {Seitzer}}]{Karachentsev2003}
{Karachentsev}, I.~D. {et~al.} 2003, \aap, 404, 93

\bibitem[{{Kim} {et~al.}(2004){Kim}, {Sung}, {Park}, \& {Sung}}]{Kim2004}
{Kim}, S.~C., {Sung}, H., {Park}, H.~S., \& {Sung}, E.-C. 2004, Chinese Journal
  of Astronomy and Astrophysics, 4, 299

\bibitem[{{Kroupa}(2001)}]{Kroupa2001}
{Kroupa}, P. 2001, \mnras, 322, 231

\bibitem[{{Kroupa}(2002)}]{Kroupa2002}
---. 2002, Science, 295, 82

\bibitem[{{Kudritzki} {et~al.}(2008){Kudritzki}, {Urbaneja}, {Bresolin},
  {Przybilla}, {Gieren}, \& {Pietrzy{\'n}ski}}]{Kudritzki2008}
{Kudritzki}, R.-P., {Urbaneja}, M.~A., {Bresolin}, F., {Przybilla}, N.,
  {Gieren}, W., \& {Pietrzy{\'n}ski}, G. 2008, \apj, 681, 269

\bibitem[{{Larson}(1976)}]{Larson1976}
{Larson}, R.~B. 1976, \mnras, 176, 31

\bibitem[{{MacArthur} {et~al.}(2009){MacArthur}, {Gonz{\'a}lez}, \&
  {Courteau}}]{MacArthur2009}
{MacArthur}, L.~A., {Gonz{\'a}lez}, J.~J., \& {Courteau}, S. 2009, \mnras, 395,
  28

\bibitem[{{Magrini} {et~al.}(2009){Magrini}, {Stanghellini}, \&
  {Villaver}}]{Magrini2009}
{Magrini}, L., {Stanghellini}, L., \& {Villaver}, E. 2009, \apj, 696, 729

\bibitem[{{Marigo} {et~al.}(2008){Marigo}, {Girardi}, {Bressan}, {Groenewegen},
  {Silva}, \& {Granato}}]{Marigo2008}
{Marigo}, P., {Girardi}, L., {Bressan}, A., {Groenewegen}, M.~A.~T., {Silva},
  L., \& {Granato}, G.~L. 2008, \aap, 482, 883

\bibitem[{{Matteucci} \& {Francois}(1989)}]{Matteucci1989}
{Matteucci}, F., \& {Francois}, P. 1989, \mnras, 239, 885

\bibitem[{{McConnachie} {et~al.}(2009){McConnachie}, {Irwin}, {Ibata},
  {Dubinski}, {Widrow}, {Martin}, {C{\^o}t{\'e}}, {Dotter}, {Navarro},
  {Ferguson}, {Puzia}, {Lewis}, {Babul}, {Barmby}, {Bienaym{\'e}}, {Chapman},
  {Cockcroft}, {Collins}, {Fardal}, {Harris}, {Huxor}, {Mackey},
  {Pe{\~n}arrubia}, {Rich}, {Richer}, {Siebert}, {Tanvir}, {Valls-Gabaud}, \&
  {Venn}}]{McConnachie2009}
{McConnachie}, A.~W. {et~al.} 2009, \nat, 461, 66

\bibitem[{{Mo} {et~al.}(1998){Mo}, {Mao}, \& {White}}]{Mo1998}
{Mo}, H.~J., {Mao}, S., \& {White}, S.~D.~M. 1998, \mnras, 295, 319

\bibitem[{{Mu{\~n}oz-Mateos} {et~al.}(2007){Mu{\~n}oz-Mateos}, {Gil de Paz},
  {Boissier}, {Zamorano}, {Jarrett}, {Gallego}, \& {Madore}}]{Munoz-Mateos2007}
{Mu{\~n}oz-Mateos}, J.~C., {Gil de Paz}, A., {Boissier}, S., {Zamorano}, J.,
  {Jarrett}, T., {Gallego}, J., \& {Madore}, B.~F. 2007, \apj, 658, 1006

\bibitem[{{Naab} \& {Ostriker}(2006)}]{Naab2006}
{Naab}, T., \& {Ostriker}, J.~P. 2006, \mnras, 366, 899

\bibitem[{{Nantais} {et~al.}(2008){Nantais}, {Huchra}, {Barmby}, \&
  {Olsen}}]{Nantais2008}
{Nantais}, J.~B., {Huchra}, J.~P., {Barmby}, P., \& {Olsen}, K.~A.~G. 2008,
  ArXiv e-prints

\bibitem[{{Olsen} {et~al.}(2004){Olsen}, {Miller}, {Suntzeff}, {Schommer}, \&
  {Bright}}]{Olsen2004}
{Olsen}, K.~A.~G., {Miller}, B.~W., {Suntzeff}, N.~B., {Schommer}, R.~A., \&
  {Bright}, J. 2004, \aj, 127, 2674

\bibitem[{{Pagel} {et~al.}(1979){Pagel}, {Edmunds}, {Blackwell}, {Chun}, \&
  {Smith}}]{Pagel1979}
{Pagel}, B.~E.~J., {Edmunds}, M.~G., {Blackwell}, D.~E., {Chun}, M.~S., \&
  {Smith}, G. 1979, \mnras, 189, 95

\bibitem[{{Pohlen} \& {Trujillo}(2006)}]{Pohlen2006}
{Pohlen}, M., \& {Trujillo}, I. 2006, \aap, 454, 759

\bibitem[{{Puche} {et~al.}(1990){Puche}, {Carignan}, \& {Bosma}}]{Puche1990}
{Puche}, D., {Carignan}, C., \& {Bosma}, A. 1990, \aj, 100, 1468

\bibitem[{{Putman} {et~al.}(2009){Putman}, {Peek}, {Muratov}, {Gnedin}, {Hsu},
  {Douglas}, {Heiles}, {Stanimirovic}, {Korpela}, \& {Gibson}}]{Putman2009}
{Putman}, M.~E. {et~al.} 2009, \apj, 703, 1486

\bibitem[{{Quillen} {et~al.}(2009){Quillen}, {Minchev}, {Bland-Hawthorn}, \&
  {Haywood}}]{Quillen2009}
{Quillen}, A.~C., {Minchev}, I., {Bland-Hawthorn}, J., \& {Haywood}, M. 2009,
  \mnras, 397, 1599

\bibitem[{{Rizzi} {et~al.}(2006){Rizzi}, {Bresolin}, {Kudritzki}, {Gieren}, \&
  {Pietrzy{\'n}ski}}]{Rizzi2006}
{Rizzi}, L., {Bresolin}, F., {Kudritzki}, R.-P., {Gieren}, W., \&
  {Pietrzy{\'n}ski}, G. 2006, \apj, 638, 766

\bibitem[{{Roussel} {et~al.}(2005){Roussel}, {Gil de Paz}, {Seibert}, {Helou},
  {Madore}, \& {Martin}}]{Roussel2005}
{Roussel}, H., {Gil de Paz}, A., {Seibert}, M., {Helou}, G., {Madore}, B.~F.,
  \& {Martin}, C. 2005, \apj, 632, 227

\bibitem[{{Ro{\v s}kar} {et~al.}(2008){Ro{\v s}kar}, {Debattista}, {Quinn},
  {Stinson}, \& {Wadsley}}]{Roskar2008a}
{Ro{\v s}kar}, R., {Debattista}, V.~P., {Quinn}, T.~R., {Stinson}, G.~S., \&
  {Wadsley}, J. 2008, \apjl, 684, L79

\bibitem[{{Sakai} {et~al.}(2004){Sakai}, {Ferrarese}, {Kennicutt}, \&
  {Saha}}]{Sakai2004}
{Sakai}, S., {Ferrarese}, L., {Kennicutt}, Jr., R.~C., \& {Saha}, A. 2004,
  \apj, 608, 42

\bibitem[{{Salpeter}(1955)}]{Salpeter1955}
{Salpeter}, E.~E. 1955, \apj, 121, 161

\bibitem[{{S{\'a}nchez-Bl{\'a}zquez} {et~al.}(2009){S{\'a}nchez-Bl{\'a}zquez},
  {Courty}, {Gibson}, \& {Brook}}]{Sanchez-Blazquez2009}
{S{\'a}nchez-Bl{\'a}zquez}, P., {Courty}, S., {Gibson}, B.~K., \& {Brook},
  C.~B. 2009, \mnras, 398, 591

\bibitem[{{Schlegel} {et~al.}(1998){Schlegel}, {Finkbeiner}, \&
  {Davis}}]{Schlegel1998}
{Schlegel}, D.~J., {Finkbeiner}, D.~P., \& {Davis}, M. 1998, \apj, 500, 525

\bibitem[{{Sellwood} \& {Binney}(2002)}]{Sellwood2002}
{Sellwood}, J.~A., \& {Binney}, J.~J. 2002, \mnras, 336, 785

\bibitem[{{Sellwood} \& {Carlberg}(1984)}]{Sellwood1984}
{Sellwood}, J.~A., \& {Carlberg}, R.~G. 1984, \apj, 282, 61

\bibitem[{{Skillman} {et~al.}(2003){Skillman}, {Tolstoy}, {Cole}, {Dolphin},
  {Saha}, {Gallagher}, {Dohm-Palmer}, \& {Mateo}}]{Skillman2003}
{Skillman}, E.~D., {Tolstoy}, E., {Cole}, A.~A., {Dolphin}, A.~E., {Saha}, A.,
  {Gallagher}, J.~S., {Dohm-Palmer}, R.~C., \& {Mateo}, M. 2003, \apj, 596, 253

\bibitem[{{Spergel} {et~al.}(2007){Spergel}, {Bean}, {Dor{\'e}}, {Nolta},
  {Bennett}, {Dunkley}, {Hinshaw}, {Jarosik}, {Komatsu}, {Page}, {Peiris},
  {Verde}, {Halpern}, {Hill}, {Kogut}, {Limon}, {Meyer}, {Odegard}, {Tucker},
  {Weiland}, {Wollack}, \& {Wright}}]{Spergel2007}
{Spergel}, D.~N. {et~al.} 2007, \apjs, 170, 377

\bibitem[{{Trujillo} \& {Pohlen}(2005)}]{Trujillo2005}
{Trujillo}, I., \& {Pohlen}, M. 2005, \apjl, 630, L17

\bibitem[{{Urbaneja} {et~al.}(2005){Urbaneja}, {Herrero}, {Bresolin},
  {Kudritzki}, {Gieren}, {Puls}, {Przybilla}, {Najarro}, \&
  {Pietrzy{\'n}ski}}]{Urbaneja2005}
{Urbaneja}, M.~A. {et~al.} 2005, \apj, 622, 862

\bibitem[{{Vila-Costas} \& {Edmunds}(1992)}]{Vila-Costas1992}
{Vila-Costas}, M.~B., \& {Edmunds}, M.~G. 1992, \mnras, 259, 121

\bibitem[{{Vlaji{\'c}} {et~al.}(2009){Vlaji{\'c}}, {Bland-Hawthorn}, \&
  {Freeman}}]{Vlajic2009}
{Vlaji{\'c}}, M., {Bland-Hawthorn}, J., \& {Freeman}, K.~C. 2009, \apj, 697,
  361

\bibitem[{{Webster} \& {Smith}(1983)}]{Webster1983}
{Webster}, B.~L., \& {Smith}, M.~G. 1983, \mnras, 204, 743

\bibitem[{{White} \& {Frenk}(1991)}]{White1991}
{White}, S.~D.~M., \& {Frenk}, C.~S. 1991, \apj, 379, 52

\bibitem[{{Wielen}(1977)}]{Wielen1977}
{Wielen}, R. 1977, \aap, 60, 263

\bibitem[{{Wielen} {et~al.}(1996){Wielen}, {Fuchs}, \& {Dettbarn}}]{Wielen1996}
{Wielen}, R., {Fuchs}, B., \& {Dettbarn}, C. 1996, \aap, 314, 438

\bibitem[{{Williams} {et~al.}(2009{\natexlab{a}}){Williams}, {Dalcanton},
  {Dolphin}, {Holtzman}, \& {Sarajedini}}]{Williams2009a}
{Williams}, B.~F., {Dalcanton}, J.~J., {Dolphin}, A.~E., {Holtzman}, J., \&
  {Sarajedini}, A. 2009{\natexlab{a}}, \apjl, 695, L15

\bibitem[{{Williams} {et~al.}(2009{\natexlab{b}}){Williams}, {Dalcanton},
  {Seth}, {Weisz}, {Dolphin}, {Skillman}, {Harris}, {Holtzman}, {Girardi}, {de
  Jong}, {Olsen}, {Cole}, {Gallart}, {Gogarten}, {Hidalgo}, {Mateo}, {Rosema},
  {Stetson}, \& {Quinn}}]{Williams2009}
{Williams}, B.~F. {et~al.} 2009{\natexlab{b}}, \aj, 137, 419

\bibitem[{{Williams} {et~al.}(2009{\natexlab{c}}){Williams}, {Dalcanton},
  {Stilp}, {Gilbert}, {Roskar}, {Seth}, {Weisz}, {Dolphin}, {Gogarten},
  {Skillman}, \& {Holtzman}}]{Williams2009b}
---. 2009{\natexlab{c}}, ArXiv e-prints

\bibitem[{{Zaritsky}(1999)}]{Zaritsky1999}
{Zaritsky}, D. 1999, \aj, 118, 2824

\bibitem[{{Zaritsky} {et~al.}(2002){Zaritsky}, {Harris}, {Thompson}, {Grebel},
  \& {Massey}}]{Zaritsky2002}
{Zaritsky}, D., {Harris}, J., {Thompson}, I.~B., {Grebel}, E.~K., \& {Massey},
  P. 2002, \aj, 123, 855

\bibitem[{{Zaritsky} {et~al.}(1994){Zaritsky}, {Kennicutt}, \&
  {Huchra}}]{Zaritsky1994}
{Zaritsky}, D., {Kennicutt}, Jr., R.~C., \& {Huchra}, J.~P. 1994, \apj, 420, 87

\end{thebibliography}

\end{document}